\documentclass[twocolumn,aps,prd,amsmath,amsfonts,nofootinbib,longbibliography]{revtex4-1}

\newcommand{\ud}{\mathrm{d}}

\newcommand{\bs}{\begin{split}}
\newcommand{\es}{\end{split}}

\newcommand{\Sh}{{(\mathrm{QB})}}
\newcommand{\FG}[2]{F_{\bar{\gamma}}^{#1}({#2})+\alpha G_{\bar{\gamma}}^{#1}({#2})}
\newcommand{\FGn}[1]{F_{\bar{\gamma}}^{#1}+\alpha G_{\bar{\gamma}}^{#1}}

\usepackage{wrapfig}
\usepackage{mathrsfs}
\usepackage{xcolor}
\usepackage{verbatim}
\usepackage{float}
\usepackage{color}
\usepackage[hyperindex,breaklinks]{hyperref}
\hypersetup{
colorlinks,
    linkcolor={black},
    citecolor={blue!50!black},
    urlcolor={blue!80!black},
    pdftitle={Quantum Bias Cosmology: Acceleration from Holographic Information Capacity},
    pdfauthor={Luke M. Butcher}}
\usepackage{slashed}
\usepackage{graphicx}
\usepackage{subfigure}
\usepackage{psfrag}
\usepackage{mathtools}
\usepackage{amsthm}
\theoremstyle{plain}

\begin{document}

\title{Quantum Bias Cosmology:\\ Acceleration from Holographic Information Capacity}
\author{Luke M. Butcher}
\email[]{lmb@roe.ac.uk}
\affiliation{Institute for Astronomy, University of Edinburgh, Royal Observatory, Edinburgh EH9 3HJ, United Kingdom}
\date{June 14, 2019}
\pacs{}

\begin{abstract}
I show that a generic quantum phenomenon can drive cosmic acceleration without the need for dark energy or modified gravity. When treating the universe as a quantum system, one typically focuses on the scale factor (of an FRW spacetime) and ignores many other degrees of freedom. However, the information capacity of the discarded variables will inevitably change as the universe expands, generating \emph{quantum bias} (QB) in the Friedmann equations. If information could be stored in each Planck-volume independently, this effect would give rise to a constant acceleration $10^{120}$ times larger than that observed, reproducing the usual cosmological constant problem. However, once information capacity is quantified according to the holographic principle, cosmic acceleration is far smaller and depends on the past behaviour of the scale factor. I calculate this holographic quantum bias, derive the semiclassical Friedmann equations, and obtain their general solution for a spatially-flat universe containing matter and radiation. Comparing these QB-CDM solutions to those of $\Lambda$CDM, the new theory is shown to be falsifiable, but nonetheless consistent with current observations. In general, realistic QB cosmologies undergo phantom acceleration ($w_\mathrm{eff}<-1$) at late times, predicting a Big Rip in the distant future.
\end{abstract} 

\maketitle

\section{introduction}
We know the universe is expanding at an accelerating rate \cite{Riess98,Perlmutter99,Peebles03,Weinberg18}, but the cause of this acceleration remains a mystery to fundamental physics \cite{Weinberg89,Carroll01,Copeland06}. Current observations are broadly consistent with the simplest proposal: acceleration driven by a  cosmological constant $\Lambda>0$ \cite{Planck18}. But if we are to understand $\Lambda$ as the energy-density of empty space, we cannot currently explain the extremely tiny value $\Lambda_{\text{obs}}\sim 10^{-122}/ \ell^{2}_\mathrm{pl} $ without anthropic reasoning \cite{Weinberg87,Douglas03,Bousso00b,Susskind03,Vilenkin06}. Alternatively, we may hope to derive cosmic acceleration from new dynamical fields, or modifications to Einstein's gravity \cite{Clifton12,Joyce15}. However, these models often struggle to fit local constraints (from the solar system \cite{Will14} and gravitational wave observations \cite{LIGO17}) and still generate the acceleration we observe \cite{Lombriser17,Baker17,Sakstein17}.

In this paper, I will motivate and develop a new explanation for cosmic acceleration -- one that does not require a cosmological constant, new dynamical fields, or modified gravity. Instead, we will examine an overlooked quantum phenomenon \cite{Butcher18a,Butcher18b} and show that its application to cosmology gives rise to a new acceleration term in the Friedmann equations. This \emph{quantum bias} depends on the maximum information the universe can hold, which we will quantify according to the \emph{holographic principle} \cite{Hooft93,Susskind95,Bousso99b,Bousso02}. Besides this step, our approach will be broadly independent of the details of quantum gravity at the fundamental level.

Empirically, this new theory has many features that distinguish it from a typical dark energy/modified gravity model. First, it describes a purely \emph{global} phenomenon: the background undergoes accelerated expansion without additional \emph{local} effects (e.g.\ perturbations in a dark fluid, or deviations from the Einstein field equations). Second, the universe can end in a Big Rip \cite{Caldwell03}, with quantum bias resembling phantom dark energy at late times. Third, the model has very little freedom: it only introduces a single new parameter, has no free functions, and cannot be tuned to mimic $\Lambda$ to arbitrary accuracy. Nonetheless, a quick comparison with $\Lambda$CDM will suggest the theory is consistent with current observations. 

We will take a systematic approach, working all the way from first principles to exact cosmological solutions. (In contrast, there are numerous attempts to link holography to dark energy that invoke \emph{ad hoc} modifications to the Friedmann equations, or derive only approximate solutions, e.g.\ \cite{Li04,Huang04, Linder04,Enqvist04,Hsu04,Wang05,Ke05,Kim06,Hu06,Almeida06,Zhang07,vanPutten15}.) Before describing how the paper will unfold, it will be helpful to first give a brief summary of the generic quantum phenomenon \cite{Butcher18a,Butcher18b} that forms the basis of this theory.

\subsection{Quantum Bias}\label{DoFs}
Suppose we are interested in an observable $x$ of some physical system with many degrees of freedom $(x,\varphi^1,\varphi^2,\ldots)$. If the classical behaviour of $x$ can be derived from an action
\begin{align}\label{reducedAction}
\mathcal{I}[x(t)]\equiv\int \ud t \left[\frac{m}{2}\dot{x}^2 -V_\mathrm{cl}(x)\right],
\end{align}
without reference to the other variables $\varphi=(\varphi^1,\varphi^2,\ldots)$, we say that the other degrees of freedom $\varphi$ can be \emph{discarded} when predicting the classical path $x(t)$. 

However, once quantum effects are considered, we cannot always continue to use the action (\ref{reducedAction}) to predict the behaviour of $x$. Indeed, if the discarded degrees of freedom have a Hilbert space $\mathcal{H}_\varphi$ that depends on $x$, with information capacity $\mathcal{S}(x)\equiv \ln(\dim[\mathcal{H}_\varphi(x)])\ne \text{const}$, then a quantum correction  will appear in the effective potential \cite{Butcher18a}:
\begin{align}\label{VofS}
\!\Delta V_\mathrm{eff}= \frac{\hbar^2}{8m}\!\left[\!\left(\!1- 4\xi \frac{d+1}{d}\!\right)\!\!\left(\partial_x \mathcal{S}\right)^2 + 2(1-4\xi)\partial_x^2 \mathcal{S}\right]\!,\!
\end{align}
where $\xi\in \mathbb{R}$ is a curvature coupling parameter, and $d\in \mathbb{N}$ the dimensionality of the discarded configuration space. (See appendix \ref{BiasDeriv} for a brief summary of the derivation of this result and a discussion of its generality.) The correction (\ref{VofS}) introduces a \emph{bias} in the behaviour of $x$:
\begin{align}\label{meanEOM}
m\partial_t^2\langle x\rangle&=- \langle \partial_x V_\mathrm{cl} + \partial_x\Delta V_\mathrm{eff}\rangle,
\end{align}
so the classical equation of motion $m\ddot{x}=-\partial_x V_\mathrm{cl} $ is no longer true on average. This motivates the use of a semiclassical action
\begin{align}\label{semiclasS}
\mathcal{J}[x(t)]\equiv\int \ud t \left[\frac{m}{2}\dot{x}^2 -V_\mathrm{cl}(x) - \Delta V_\mathrm{eff}(x)\right],
\end{align}
which generates trajectories consistent with the average motion (\ref{meanEOM}). Moreover, the semiclassical action (\ref{semiclasS}) sets the phase of paths $x(t)$ in the path integral, once the discarded variables have been integrated out \cite{Butcher18b}.

\subsection{Outline of Paper}
The aim of this article is to apply the above results to cosmology. The universe is clearly a quantum system with many degrees of freedom;\footnote{The laws of quantum mechanics are expected to apply to all physical systems, and the universe is no exception. The question is: how accurate is the \emph{classical approximation} to the universe that we typically use in cosmology? In general, this approximation will only be accurate when quantum bias (\ref{VofS}) can be neglected.} moreover, the classical behaviour of its scale factor $a$ can be derived from an action of the form (\ref{reducedAction}). Hence, if the other degrees of freedom have an information capacity $\mathcal{S}(a)\ne\text{const}$, we should expect there to be a quantum bias (\ref{VofS}) forcing $a(t)$ off its classical trajectory. We wish to determine whether this effect can explain the cosmic acceleration we observe today.

The paper will proceed as follows. In section \ref{ClassAct}, we construct an action similar to (\ref{reducedAction}) that generates the classical behaviour of the scale factor $a(t)$ of an FRW spacetime. In section  \ref{QuantCor}, we obtain the quantum bias (\ref{VofS}) from the other degrees of freedom, with information capacity fixed according to the holographic principle. In section \ref{Deriv}, having assembled the semiclassical action (\ref{semiclasS}), we derive the semiclassical Friedmann equations. In section \ref{Solutions}, we solve these equations for a spatially flat universe containing matter and radiation. Finally, in section \ref{Compare}, we compare these solutions to $\Lambda$CDM, and argue that the new theory is likely to be consistent with current observations.

\section{Classical Action}\label{ClassAct}
Here we lay out our basic definitions and derive the action (\ref{reducedAction}) that encodes classical cosmology. It is important to realise that we cannot simply write down an action $\mathcal{I}[a(t)]$ and check that it generates the classical Friedmann equations. We must also ensure that the \emph{normalisation} of the action is correct, as this is critical for quantum effects. Hence we work from first principles, starting with the action for general relativity:
\begin{align}\label{IG+IM}
\mathcal{I} &= \mathcal{I}_\mathrm{G}[g_{\mu\nu}] + \mathcal{I}_\mathrm{M}[g_{\mu\nu},\Psi],
\\\label{IG}
\mathcal{I}_\mathrm{G}&\equiv \frac{1}{2\kappa} \int_\mathcal{M} \!\!\!\ud^4x\sqrt{-g}\,R  +\frac{1}{\kappa} \int_\mathcal{\partial M}\!\!\!\!\!\! \ud^3 y\, \epsilon\sqrt{|h|}\,K,
\end{align}
where the Gibbons--Hawking--York term \cite{York72,Gibbons77,Hawking95} is included for regions $\mathcal{M}$ with nontrivial boundary $\partial\mathcal{M}\ne \emptyset$.\footnote{We set $c=1$, write $\kappa\equiv 8\pi G$, $g\equiv \det(g_{\mu\nu})$, $h\equiv \det(h_{\mu\nu})$,  and adopt the sign conventions of Wald \cite{Wald}: $\eta_{\mu\nu}\equiv \mathrm{diag}(-1,1,1,1)$, $[\nabla_\mu,\nabla_\nu]v^\alpha\equiv R^{\alpha}_{\phantom{\alpha}\beta\mu\nu}v^\beta$, $R_{\mu\nu}\equiv R^{\alpha}_{\phantom{\alpha}\mu\alpha\nu}$. The metric $h_{\mu\nu}\equiv g_{\mu\nu}-\epsilon n_{\mu}n_{\nu}$ and extrinsic curvature $K_{\mu\nu}\equiv h_\mu{}^\alpha\nabla_\alpha n_\nu$ of the boundary  $\partial \mathcal{M}$ are constructed from the outward unit normal $n^\mu$, with $\epsilon \equiv n^\alpha n_\alpha =\pm 1$.} We use the generic symbol $\Psi$ to denote matter, having energy-momentum tensor 
\begin{align}\label{Tdef}
T_{\mu\nu}\equiv \frac{-2}{\sqrt{-g}}\frac{\delta \mathcal{I}_\mathrm{M}}{\delta g^{\mu\nu}},
\end{align}
and set the cosmological constant $\Lambda=0$, the aim being to generate cosmic acceleration nonetheless.

\subsection{FRW Spacetime}
To construct an action of the form (\ref{reducedAction}) we must discard almost all the degrees of freedom in general relativity, restricting the action (\ref{IG}) to spacetimes that are completely homogeneous and isotropic. It is convenient to use the following form of the FRW metric:
\begin{align}\label{FRWmetric}
\ud s^2 = [a(t)]^2\left(-[N(t)]^2\ud t^2 + \ud \chi^2 + [r_k(\chi)]^2\ud \Omega^2\right),
\end{align}
where $a(t)$ is the scale factor, $\chi$ the comoving distance, and $\ud \Omega^2=\ud\theta^2 +\sin^2\!\theta\,\ud \phi^2$. The lapse function $N(t)$ controls the gauge of the time coordinate $t$,\footnote{We cannot fix the gauge at this stage because we will need to take variations $\delta N(t)$, in addition to $\delta a(t)$, to obtain \emph{both} Friedmann equations from the action (\ref{IG}). Afterwards, we will adopt the gauge $N(t)=1$ in which $t$ is equivalent to  conformal time $\eta$.} and the spatial geometry is described by the function
\begin{align}
r_k(\chi)&\equiv
\begin{cases}
\sin(\chi), &k=+1,\\
\chi, &k=0,\\
\sinh(\chi), &k=-1,
\end{cases}
\end{align}
for a closed, flat, or open universe respectively. (Note that $\chi$ is dimensionless, and $a$ is the radius of spatial curvature for $k\ne0$.) As such, a surface of constant $\chi$ and $t$ is a sphere of area $A=\mathscr{A}(\chi)[a(t)]^2 $ and volume $V= \mathscr{V}(\chi)[a(t)]^3$, where
\begin{align}\label{AVdef}
\mathscr{A}(\chi)&\equiv 4\pi [r_k(\chi)]^2 ,& \mathscr{V}(\chi)&\equiv 4 \pi  \int^{\chi}_0\! \ud \chi'[r_k(\chi')]^2.
\end{align}
For the sake of evaluating $\mathcal{I}_\mathrm{G}$, we will also need the scalar curvature of the FRW spacetime (\ref{FRWmetric}):
\begin{align}\label{R}
R= \frac{6}{a^2 N^2}\left(\frac{\ddot{a}}{a}- \frac{\dot{a}\dot{N}}{a N}  + k N^2\right),
\end{align}
where dots indicate differentiation with respect to $t$.

\subsection{Integration Region and Boundary}
Besides evaluating the action (\ref{IG}) on the metric (\ref{FRWmetric}), we must also choose a suitable region $\mathcal{M}$ over which to integrate. Rather than attempt an integral over all space (with an infinite result for $k\in\{0,-1\}$) we limit ourselves to the spherically symmetric region
\begin{align}\label{Mdef}
\mathcal{M}:\quad 
\begin{array}{rlrl}
t\!\! &\in [t_-,t_+],&\quad\theta\!\!&\in [0,\pi],\\
\chi\!\!&\in [0,\chi_*],&\quad \phi \!\!&\in [0,2\pi),
\end{array}
\end{align} 
and promise to send $\chi_*\to \infty$ (or $\chi_*\to \pi$, for $k=1$) at the end of the calculation. It is easy to see that the boundary of (\ref{Mdef}) has three components: $\partial\mathcal{M}=\partial\mathcal{M}_{\chi_*}\cup \partial\mathcal{M}_{t_-}\cup \partial\mathcal{M}_{t_+}$; their extrinsic scalar curvatures are
\begin{align}\label{Ks}
K[\partial\mathcal{M}_{\chi_*}]&= \frac{\mathscr{A}'_{*}}{\mathscr{A}_* a},& K[\partial\mathcal{M}_{t_\pm}]&=\left.\pm\frac{3 \dot{a}}{a^2N}\right|_{t=t_\pm},
\end{align}
where the prime denotes a derivative, and asterisks indicate evaluation at $\chi=\chi_*$. With $\mathcal{M}$ defined, we can now discuss the matter action $\mathcal{I}_\mathrm{M}$, and then evaluate the gravitational action $\mathcal{I}_\mathrm{G}$ on the FRW metric (\ref{FRWmetric}).

\subsection{Matter Action}
In order to provide matter terms for the Friedmann equations, we require formulae for the functional derivatives of $\mathcal{I}_\mathrm{M}$ with respect to variations $\delta a(t)$, $\delta N(t)$ in the FRW metric (\ref{FRWmetric}). Note that these variations cause the inverse metric to change by
\begin{align}\label{deltag}
\delta g^{\mu\nu}= -2 g^{\mu\nu}\frac{\delta a}{a} - 2 g^{00}\delta^\mu_{0}\delta^\nu_{0} \frac{\delta N}{N},
\end{align}
and hence the matter action varies according to
\begin{align}\nonumber
\delta \mathcal{I}_\mathrm{M}&= \int_\mathcal{M} \ud^4x\, \frac{\delta \mathcal{I}_\mathrm{M}}{\delta g^{\mu\nu}} \delta g^{\mu\nu}\\ \nonumber
&= \int_\mathcal{M} \ud^4x\, \frac{\sqrt{-g}}{(-2)} T_{\mu\nu}\left[-2 g^{\mu\nu}\frac{\delta a}{a} - 2 g^{00}\delta^\mu_{0}\delta^\nu_{0} \frac{\delta N}{N}\right]\\\label{deltaIM}
&= \int_\mathcal{M} \ud^4x\, \sqrt{-g}\left[T\frac{\delta a}{a} + T_{00}g^{00}\frac{\delta N}{N}\right],
\end{align}
where we used (\ref{Tdef}) in the second line. Homogeneous and isotropic matter $\Psi=\Psi(t)$ has energy-density $\rho=\rho(t)$ and pressure $p=p(t)$ that depend on $t$ only, with $T=3p -\rho$ and $T_{00}g^{00}=-\rho$. As such, equation (\ref{deltaIM}) becomes
\begin{align}
\delta \mathcal{I}_\mathrm{M}&= \mathscr{V}_* \int_{t_-}^{t_+} \ud t\, a^3\left[N (3p-\rho) \delta a - a \rho \delta N \right].
\end{align}
Consequently,
\begin{align}\label{dIM}
\frac{\delta \mathcal{I}_\mathrm{M}}{\delta a}&= \mathscr{V}_* a^3 N \left(3p -\rho\right),& \frac{\delta \mathcal{I}_\mathrm{M}}{\delta N}&= -\mathscr{V}_*  a^4 \rho,
\end{align}
are the functional derivatives we need.

\subsection{Gravitational Action}
Finally, we assemble the gravitational part of the classical action by inserting (\ref{R}) and (\ref{Ks}) into (\ref{IG}). After integrating the $\ddot{a}$ term by parts (to cancel the contributions from $\partial\mathcal{M}_{t_\pm}$) we obtain
\begin{align}\label{IGboundary}
\mathcal{I}_\mathrm{G}= \frac{3\mathscr{V}_*}{\kappa}\!\int^{t_+}_{t_-}\!\! \ud t \!\left[ -\frac{\dot{a}^2}{N} + k N a^2  \right] + \frac{\mathscr{A}'_*}{\kappa}\!\int^{t_+}_{t_-}\!\! \ud t\, Na^2.
\end{align}
In general, the integral proportional to $\mathscr{A}'_*$ can be dropped when $\mathcal{M}$ covers the entire space. For $k=0$, this happens in the obvious fashion: $\mathscr{V}_*=4\pi \chi_*^3/3 $ and $\mathscr{A}'_*=8 \pi \chi_*$, so the first integral dominates over the second in the limit $\chi_*\to \infty$. For $k=1$, the full space is covered by sending $\chi_*\to\pi$, with $\mathscr{V}_*\to2\pi^2$ and $\mathscr{A}'_*\to0$ as a result. Thus, the full-space limit gives
\begin{align}\label{IGaN}
\mathcal{I}_\mathrm{G}[a(t),N(t)]= \frac{3\mathscr{V}_*}{\kappa}\int^{t_+}_{t_-} \ud t \left[- \frac{\dot{a}^2}{N} + k N a^2  \right],
\end{align}
for $k\in \{0,1\}$ at least.\footnote{For $k=-1$, $\mathscr{A}'_*\sim 4\mathscr{V}_*$ as $\chi_*\to\infty$, so (\ref{IGaN}) cannot be obtained from the limit of (\ref{IGboundary}).}  This fixes the normalisation of the total action (\ref{IG+IM}), being the sum of  the gravitational action (\ref{IGaN}) and a matter action $\mathcal{I}_\mathrm{M}$ with derivatives (\ref{dIM}). It is easy to check that this combination generates the correct Friedmann equations for the metric (\ref{FRWmetric}). Moreover, these equations are correct for all $k\in \{-1,0,1\}$, so (\ref{IGaN}) must be the correctly normalised classical action, even for an open universe.

To complete our calculation, we express (\ref{IGaN}) in terms of the conformal time coordinate $\eta=\eta(t)$, defined by
\begin{align}\label{etadef}
\ud \eta&=  N \ud t, & \eta_\pm&\equiv \eta(t_\pm),
\end{align}
and find that $N$ drops out completely: 
\begin{align}\label{IGa}
\mathcal{I}_\mathrm{G}[a(\eta)]= \frac{3\mathscr{V}_*}{\kappa}\int^{\eta_+}_{\eta_-} \ud \eta \left[ -\left(\frac{\ud{a}}{\ud\eta}\right)^2 + k a^2 \right].
\end{align}
This classical action has exactly the form (\ref{reducedAction}) we require.

\section{Cosmological Quantum Bias}\label{QuantCor}
To calculate the cosmological effect of quantum bias, we first compare the classical action (\ref{IGa}) to the standard form (\ref{reducedAction}): formally identifying $x\to a$, $t\to \eta$, and $m\to-6\mathscr{V}_*/\kappa$, the quantum bias (\ref{VofS}) becomes
\begin{align}\nonumber
\Delta V_\mathrm{eff}&= -\frac{4 \pi^2 \ell_\mathrm{pl}^4}{3 \mathscr{V}_*\kappa}\left[\left(1- 4\xi \frac{d+1}{d}\right)\left(\partial_a \mathcal{S}\right)^2\right. \\\label{VofSa}
&\quad{} + 2(1-4\xi)\partial_a^2 \mathcal{S}\bigg],
\end{align}
where  $\ell_\mathrm{pl}\equiv \sqrt{\hbar\kappa/8\pi}$ is the Planck length.\footnote{As covered in appendix \ref{BiasDeriv}, the path integral derivation of $\Delta V_\mathrm{eff}$ ensures that (\ref{VofSa}) is valid for the general form $\mathcal{S}=\mathcal{S}\!\left(a(\eta),\int^\eta \ud \eta' f(a(\eta'))\right)$, with $\partial_a$ derivatives acting on the first argument of $\mathcal{S}$ only \cite{Butcher18b}. This includes the case $\mathcal{S}=\mathcal{S}(a(\eta),\eta)$ that will be most useful here.} The bias $\Delta V_\mathrm{eff}$ arises from the many quantum degrees of freedom we have discarded by describing the universe in terms of the single observable $a(\eta)$ -- all the particles and inhomogeneities that could exist within the spatial region $\chi\in[0,\chi_*]$. Although we would need a complete understanding of quantum gravity to describe these fundamental degrees of freedom in detail, the holographic principle will suffice to fix their maximum entropy/information $\mathcal{S}$; we can then treat $\xi$ and $d$ as unknown constants, to be determined by experiment.

I now claim that we can drop the $\partial_a^2 \mathcal{S}$ term in (\ref{VofSa}) and simply write
\begin{align}\label{dropddS}
\Delta V_\mathrm{eff}=\frac{4 \pi^2 \ell_\mathrm{pl}^4}{3 \mathscr{V}_*\kappa}\left(4\xi \frac{d+1}{d}-1\right)\left(\partial_a \mathcal{S}\right)^2.
\end{align}
There are two distinct reasons for this.  The first is practical: $(\partial_a\mathcal{S})^2\sim \mathcal{S}^2/a^2$ is far bigger than $\partial_a^2 \mathcal{S}\sim \mathcal{S}/a^2$ whenever the information capacity is very large, i.e.\ $\mathcal{S}\gg1$. This will always be the case for regions $\chi\in[0,\chi_*]$ that are much larger than the Planck length: $a\chi_*\gg \ell_\mathrm{pl}$. We can take this for granted as $\chi_*\to \infty$ for $k\in\{0,-1\}$; for $k=1$, it can only fail if the universe is Planckian ($a\pi \sim \ell_\mathrm{pl}$) and therefore unsuitable for a semiclassical treatment anyway.

The second reason is theoretical: even though the $\partial_a^2\mathcal{S}$ contribution is tiny, it is not exactly zero, so it retains the potential to break a symmetry of the classical theory. In appendix \ref{Gauge}, I show that this is indeed the case. The classical theory has a gauge freedom $N(t)$, and is also invariant under a redefinition of the dynamical variable $a\to a(\widetilde{a}(t))$; it turns out that the $\partial_a^2\mathcal{S}$ term breaks this combined symmetry. Therefore, to insist that $\Delta V_\mathrm{eff}$ respect both these classical symmetries compels us to set $\xi =1/4$ and banish the $\partial_a^2\mathcal{S}$ term entirely. The result is equation (\ref{dropddS}) with the replacement
\begin{align}
\left(4\xi \frac{d+1}{d}-1\right)\to\frac{1}{d}.
\end{align}
Given that we cannot properly interpret $\xi$ or $d$ without reference to a theory of quantum gravity, it seems wise to retain the full generality of $\xi \in \mathbb{R}$, despite this symmetry argument. Nonetheless, this discussion motivates us to absorb $\xi$ and $d$ into a single dimensionless parameter
\begin{align}\label{dbardef}
\bar{d} \equiv \left(4\xi \frac{d+1}{d}-1\right)^{-1}\in \mathbb{R},
\end{align}
so that (\ref{dropddS}) becomes
\begin{align}\label{Vsimple}
\Delta V_\mathrm{eff}&= \frac{4 \pi^2 \ell_\mathrm{pl}^4}{3  \mathscr{V}_*\kappa \bar{d}}\left(\partial_a \mathcal{S}\right)^2,
\end{align}
with  $\bar{d}=d$ for the symmetric case $\xi=1/4$. As such, the symmetry argument restricts $\bar{d}\in\mathbb{N}$ for the minimal model of discarded degrees of freedom (\ref{DoFmetric}), while the generalisation (\ref{DoFmetricShape}) allows $\bar{d}\in\mathbb{R}^+$. In general, we will use an overbar to label the key dimensionless parameters of the theory.

\subsection{Volumetric Information Capacity}
Before we invoke the holographic principle, it is instructive to first consider a counterfactual argument, based on the naive idea that one should be able to store information in every Planck volume independently.  This discussion will connect our work to the old cosmological constant problem, and serve as a warm up for the holographic calculation to come.

So suppose it were possible to store exactly $n$ qubits in every Planck volume. Then the information capacity of the region $\chi \in [0,\chi_*]$ would be
\begin{align}\label{Svol}
\mathcal{S}_\mathrm{v}=  n \ln 2\cdot \mathscr{V}_* a^3/\ell_\mathrm{pl}^3,
\end{align}
leading to a quantum bias (\ref{Vsimple}) as follows:
\begin{align}\label{Vvol}
\Delta V_\mathrm{eff}&= \frac{12 \pi^2   n^2 (\ln 2)^2 \mathscr{V}_* a^4}{ \kappa \ell_\mathrm{pl}^2 \bar{d}}.
\end{align}
We would then construct the semiclassical action (\ref{semiclasS}) by inserting  quantum bias (\ref{Vvol}) into the classical action (\ref{IGa}):
 \begin{align}\label{Jvol}
\mathcal{J}_\mathrm{G}&= \mathcal{I}_\mathrm{G}[a(\eta)] -  \int^{\eta_+}_{\eta_-} \ud \eta\, \Delta V_\mathrm{eff}\\ \nonumber
&=  \frac{3\mathscr{V}_*}{\kappa}\int^{\eta_+}_{\eta_-} \ud \eta \left[ -\left(\frac{\ud{a}}{\ud\eta}\right)^2 + k a^2  -\frac{4 \pi^2  n^2 (\ln 2)^2 a^4}{\ell_\mathrm{pl}^2 \bar{d}}\right]\\ \nonumber
&= \frac{3\mathscr{V}_*}{\kappa}\int^{t_+}_{t_-} \ud t \left[-\frac{\dot{a}^2}{N} + k N a^2 -\frac{4 \pi^2  n^2 (\ln 2)^2 Na^4}{\ell_\mathrm{pl}^2 \bar{d}}\right].
\end{align}
But notice: the quantum bias term closely resembles the contribution from a cosmological constant,
\begin{align}
\mathcal{I}_\Lambda=-\frac{1}{\kappa}\int_\mathcal{M} \ud^4 x \sqrt{-g}\,\Lambda = -\frac{\mathscr{V}_*}{\kappa} \int^{t_+}_{t_-} \ud t\, Na^4 \Lambda.
\end{align}
In other words, the semiclassical action (\ref{Jvol}) is 
\begin{align}
\mathcal{J}_\mathrm{G}&= \frac{3\mathscr{V}_*}{\kappa}\int^{t_+}_{t_-} \ud t \left[ -\frac{\dot{a}^2}{N} + k N a^2 -\frac{\Lambda_\mathrm{eff} N a^4}{3}\right],
\end{align}
with an effective cosmological constant 
\begin{align}\label{Lambdaeff}
\Lambda_\mathrm{eff}&= \frac{12 \pi^2  n^2 (\ln 2)^2}{\ell_\mathrm{pl}^2 \bar{d}}.
\end{align}
For $n,\bar{d}\sim 1$, we see that $\Lambda_\mathrm{eff}\sim 10^{124}\Lambda_\mathrm{obs}$ reproduces the enormous cosmological constant that normally arises from summing zero-point energies up to the Planck scale.

\emph{A priori}, there was no reason to expect a connection between cosmological quantum bias (\ref{VofSa}) and vacuum energy. Nonetheless, when we place independent degrees of freedom in each Planck volume (\ref{Svol}) these two phenomena generate the same cosmic acceleration (\ref{Lambdaeff}). It is unclear whether this resemblance is purely superficial, or evidence of some fundamental connection between vacuum energy and quantum bias. The second option suggests an exciting possibility: counting degrees of freedom correctly (i.e.\ holographically) may not only suffice to generate the cosmic acceleration we do observe, but could also \emph{explain away} the large vacuum energy predicted by quantum field theory. We leave this discussion for another time, content to tackle the former problem, without a definitive answer to the latter.

\subsection{Holographic Information Capacity}\label{HoloCap}
In fact, the volumetric formula (\ref{Svol}) is wrong: information cannot be stored in each Planck volume independently. As detailed in appendix \ref{HoloPrinc}, quantum gravity considerations (the holographic principle \cite{Hooft93, Susskind95, Bousso99b, Bousso02} and black hole complementarity \cite{Susskind93,Susskind94}) lead us instead to the following formula for  the information capacity of the region $\chi \in [0,\chi_*]$ at conformal time $\eta$:
\begin{align}\label{Sholo}
\mathcal{S}_\mathrm{h}(a,\eta)=  \frac{\mathscr{A}(\bar{\eta}-\eta) a^2}{4 \ell^2_\mathrm{pl}} \cdot\frac{\bar{\mu} \mathscr{V}_*}{\mathscr{V}(\bar{\eta}-\eta)},
\end{align}
where $\bar{\eta}$ is the \emph{final conformal time} (the limiting value of $\eta$ in the far future) $\bar{\mu}=1/(24 \ln 2 -15) \approx 0.61142$ is a numerical constant, and the functions $\mathscr{A}(\cdot)$ and $\mathscr{V}(\cdot)$ measure the comoving area and volume of a sphere (\ref{AVdef}). In equation (\ref{Sholo}) the first fraction quantifies the information capacity of a sphere the size of the cosmological event horizon, and the second fraction is the number of these spheres inside $\chi \in [0,\chi_*]$. (The filling factor $\bar{\mu}$ accounts for the organisation of holographic information in spacetime; see appendix \ref{HoloPrinc} for details.) In section \ref{etabarcheck}, we will confirm that $\bar{\eta}$ really is the final conformal time: quantum bias $\Delta V_\mathrm{eff}$ generates cosmic acceleration that inevitably sends $a(\eta)\to \infty$ as $\eta \to \bar{\eta}$.

The derivation of (\ref{Sholo}) assumes that the universe is expanding $\dot{a}>0$, and that the event horizon is far smaller than the radius of spatial curvature: $|k|(\bar{\eta}-\eta)\ll 1$. For our universe, these assumptions can only break down at very early times, either during inflation, or before a Big Bounce. Hence, equation (\ref{Sholo}) is certainly suitable for a theory of late-time cosmic acceleration. (I will  revisit these assumptions in a future publication, when I examine the role of quantum bias in the very early universe.) At the very least, a reader who is sceptical of the arguments in appendix \ref{HoloPrinc} can always take (\ref{Sholo}) to be a well-motivated \emph{holographic hypothesis}, the cosmological consequences of which we will now examine in detail.

We begin, as with the volumetric case, by calculating the quantum bias (\ref{Vsimple}):
\begin{align}\label{Vholo}
\Delta V_\mathrm{eff}&= \frac{ 3\pi^2 \bar{\mu}^2 \mathscr{V}_*}{\kappa \bar{d}}\left(\frac{ \mathscr{A}(\bar{\eta}-\eta)}{3\mathscr{V}(\bar{\eta}-\eta)}\right)^2 a^2.
\end{align}
Once again, this combines with the classical action (\ref{IGa}) to form the semiclassical action (\ref{semiclasS}):
\begin{align}\nonumber
\mathcal{J}_\mathrm{G}[a(\eta)]&= \mathcal{I}_\mathrm{G}[a(\eta)] -  \int^{\eta_+}_{\eta_-} \ud \eta\, \Delta V_\mathrm{eff}\\ \nonumber
&=  \frac{3\mathscr{V}_*}{\kappa}\int^{\eta_+}_{\eta_-} \ud \eta \Bigg[ -\left(\frac{\ud{a}}{\ud\eta}\right)^2 + k a^2 \\\label{JGa}
&\quad {} - \frac{ \pi^2 \bar{\mu}^2}{ \bar{d} }\left(\frac{ \mathscr{A}(\bar{\eta}-\eta)}{3\mathscr{V}(\bar{\eta}-\eta)}\right)^2a^2\Bigg].
\end{align}
Notice that the integration limits $\eta_\pm$ determine the interval over which this action defines the dynamics of the spacetime. There is no reason to truncate our theory at late times, so we must send $\eta_+ \to \bar{\eta}$. On the other hand, we may want to keep $\eta_-$ as a cutoff at early times, for when energy-densities approach the Planck-scale and the semiclassical approximation breaks down. In general, the details of this Planckian cutoff $\eta_-\sim \eta_\mathrm{pl}$ will only be relevant at very early times; after the end of inflation, we can model the universe as containing only matter and radiation, and conflate the cutoff with the classical Big Bang: $a(\eta_-)=0$.

Finally, we re-express the semiclassical action (\ref{JGa}) in terms of the generic time coordinate $t$, so that we have two dynamical variables $(a, N)$ with which to derive the two semiclassical Friedmann equations. Recalling the definition of conformal time (\ref{etadef}) the action (\ref{JGa}) becomes
\begin{align}\nonumber
\mathcal{J}_\mathrm{G}[a(t),N(t)]&= \frac{3\mathscr{V}_*}{\kappa}\int^{\bar{t}}_{t_-} \ud t \left[-\frac{\dot{a}^2}{N} + k N a^2  \right.
 \\\label{JGaN}
 &\quad \left. {}-\left(\frac{ \mathscr{A}\big(\int^{\bar{t}}_{t} N(t') \ud t'\big)}{3\mathscr{V}\big(\int^{\bar{t}}_{t} N(t') \ud t'\big)}\right)^2 \!\bar{g} N a^2\!\right]\!,
\end{align}
where 
\begin{align}
\bar{g} &\equiv \frac{\pi^2 \bar{\mu}^2}{\bar{d}}
\end{align}
is a convenient shorthand, and $\bar{t} \in \mathbb{R}\cup \{\infty\}$ is the final value of the $t$ coordinate:
\begin{align}
\lim_{t \to \bar{t}} \eta(t)=\bar{\eta}.
\end{align}
The semiclassical action (\ref{JGaN}) is the first major result of this paper. Even though this action includes an unusual ``integral inside the integral'' term, it will still define well-behaved equations of motion. These are obtained in the next section, by infinitesimal variations $\delta a(t)$ and $\delta N(t)$.

\section{Semiclassical Friedmann Equations}\label{Deriv}
The semiclassical Friedmann equations are the equations of motion generated by the \emph{total} semiclassical action, comprising both gravitational and matter parts:
\begin{align}\label{Jdef}
\mathcal{J}\equiv \mathcal{J}_\mathrm{G} + \mathcal{I}_\mathrm{M}.
\end{align}
 (It is purely by convention that we absorb cosmological quantum bias (\ref{Vholo}) into the gravitational action; really, it is a correction to the total action: $\mathcal{I}\to \mathcal{J}$.) As usual, these equations follow by insisting that $\delta\mathcal{J}=0$ under arbitrary infinitesimal variations $\delta a(t)$, $\delta N(t)$ in the trajectories $a(t)$, $N(t)$. Given that functional derivatives of the matter action (\ref{dIM})  are already known, our main task is to obtain the derivatives $\delta \mathcal{J}_\mathrm{G}/\delta a(t)$ and $\delta \mathcal{J}_\mathrm{G}/\delta N(t)$.

\subsection{Functional Derivatives}
Rather than proceed directly from the general formula (\ref{JGaN}) we first  recall the assumption $|k|(\bar{\eta}-\eta)\ll 1$, and hence use the series expansion
\begin{align}
\left(\frac{\mathscr{A}(\chi)}{3\mathscr{V}(\chi)}\right)^2= \frac{1}{\chi^2} - \frac{4 k}{15} + O\!\left(|k|\chi^2\right)
\end{align}
to neglect terms $O(|k|(\bar{\eta}-\eta)^2)$ in the action (\ref{JGaN}):
\vfill
\begin{widetext}
\begin{align}\label{JGaNapprox}
\mathcal{J}_\mathrm{G}&= \frac{3\mathscr{V}_*}{\kappa}\int^{\bar{t}}_{t_-} \ud t \left[-\frac{\dot{a}^2}{N} +\left(1 +  \frac{4  \bar{g}}{15 } \right)kN a^2 
 -\frac{ \bar{g}N a^2}{\left(\int^{\bar{t}}_{t} N(t') \ud t'\right)^{2}}\right].
\end{align}
It is straightforward to take the functional derivative of this action with respect to the scale factor:
\begin{align}\label{dJGa}
\frac{\delta \mathcal{J}_\mathrm{G}}{\delta a(t)}&= \frac{6\mathscr{V}_*}{\kappa} \left[ \frac{\ud}{\ud t}\left(\frac{\dot{a}}{N}\right) + \left(1 +  \frac{4\bar{g}}{15 } \right)kN a 
-  \frac{\bar{g}N a}{\left(\int^{\bar{t}}_{t} N(t') \ud t'\right)^{2}}\right].
\end{align}
However, the $N(t)$ derivative requires a little more care. Under a variation $\delta N(t)$, the action (\ref{JGaNapprox}) changes by
\begin{align} \label{deltaJG1}
\delta \mathcal{J}_G &=  \frac{3\mathscr{V}_*}{\kappa}\int^{\bar{t}}_{t_-} \ud t\!\left[\delta N(t)\left(\frac{\dot{a}^2}{N^2} + \left(1 +  \frac{4\bar{g}}{15} \right) k  a^2   -   \frac{\bar{g} a^2}{\left(\int^{\bar{t}}_{t} N(t') \ud t'\right)^{2}}\right)
 +  \frac{2\bar{g}Na^2}{\left(\int^{\bar{t}}_{t} N(t') \ud t'\right)^{3}} \int^{\bar{t}}_{t} \delta N(t'') \ud t''\right].
\end{align}
We can then swap the order of integration in the last term:
\begin{align}
 \int^{\bar{t}}_{t_-} \ud t \left[\frac{N(t)[a(t)]^2}{\left(\int^{\bar{t}}_{t} N(t') \ud t'\right)^{3}} \int^{\bar{t}}_{t} \delta N(t'') \ud t'' \right]
=\int^{\bar{t}}_{t_-} \ud t \int^{\bar{t}}_{t} \ud t'' \frac{N(t)[a(t)]^2  \delta N(t'')}{\left(\int^{\bar{t}}_{t} N(t') \ud t'\right)^{3}} 
=\int^{\bar{t}}_{t_-} \ud t'' \int^{t''}_{t_-} \ud t \frac{N(t)[a(t)]^2  \delta N(t'')}{\left(\int^{\bar{t}}_{t} N(t') \ud t'\right)^{3}},
\end{align}
which becomes
\begin{align}
\int^{\bar{t}}_{t_-} \ud t'' \, \delta N(t'') \left[\int^{t''}_{t_-} \ud t \frac{N(t)[a(t)]^2 }{\left(\int^{\bar{t}}_{t} N(t') \ud t'\right)^{3}} \right]
=\int^{\bar{t}}_{t_-} \ud t\, \delta N(t) \left[\int^{t}_{t_-} \ud t'' \frac{N(t'')[a(t'')]^2}{\left(\int^{\bar{t}}_{t''} N(t') \ud t'\right)^{3}}\right],
\end{align}
after relabelling the dummy variables $t\leftrightarrow t''$. Hence, equation (\ref{deltaJG1}) is equivalent to
\begin{align}\
\delta \mathcal{J}_G &=  \frac{3\mathscr{V}_*}{\kappa}\int^{\bar{t}}_{t_-} \ud t\,\delta N(t)\left[\frac{\dot{a}^2}{N^2} + \left(1 +  \frac{4\bar{g}}{15 } \right) k  a^2 
- \frac{\bar{g}a^2}{\left(\int^{\bar{t}}_{t} N(t') \ud t'\right)^{2}} 
+ 2\bar{g}\int^{t}_{t_-} \ud t''    \frac{N(t'') [a(t'')]^2}{\left(\int^{\bar{t}}_{t''} N(t') \ud t'\right)^{3}}\right],
\end{align}
which implies
\begin{align}\label{dJGN}
\frac{\delta \mathcal{J}_G}{\delta N(t)} &=   \frac{3\mathscr{V}_*}{\kappa}\left[\frac{\dot{a}^2}{N^2} + \left(1 +  \frac{4\bar{g}}{15 } \right) k  a^2 
 -   \frac{\bar{g}a^2}{\left(\int^{\bar{t}}_{t} N(t') \ud t'\right)^{2}}
  + 2\bar{g}\int^{t}_{t_-} \ud t''    \frac{N(t'') [a(t'')]^2}{\left(\int^{\bar{t}}_{t''} N(t') \ud t'\right)^{3}}\right].
\end{align}

\subsection{Results}\label{FEsec}
We now have all we need to assemble the semiclassical Friedmann equations. Combining equations (\ref{dIM}), (\ref{dJGa}) and (\ref{dJGN}), we see that the total semiclassical action (\ref{Jdef}) is stationary if and only if 
\begin{subequations}\label{FEgen}
\begin{align}\label{FEgen1}
 \frac{\dot{a}^2}{N^2} &= \frac{\kappa }{3} \rho   a^4
 - \left(1 +  \frac{4\bar{g}}{15  } \right) k  a^2
+   \frac{\bar{g} a^2}{\left(\int^{\bar{t}}_{t} N(t') \ud t'\right)^{2}}
- 2\bar{g} \int^{t}_{t_-} \ud t''    \frac{N(t'') [a(t'')]^2}{\left(\int^{\bar{t}}_{t''} N(t') \ud t'\right)^{3}},
\\\label{FEgen2}
\frac{\ud}{\ud t}\left(\frac{\dot{a}}{N}\right)&=\frac{\kappa}{6}  \left(\rho-3p \right)a^3 N 
 - \left(1 +  \frac{4\bar{g}}{15} \right)kN a  
 +  \frac{\bar{g}N a}{\left(\int^{\bar{t}}_{t} N(t') \ud t'\right)^{2}} .
\end{align}
\end{subequations}
Note that $\mathscr{V}_*$ has dropped out of these equations, so we are now free to send $\chi_*\to \infty$ as desired. Differentiating (\ref{FEgen1}) with respect to $t$, and comparing the result with (\ref{FEgen2}), we see that the two equations are indeed consistent, provided matter obeys the standard continuity equation:
\begin{align}
a\dot{\rho} + 3\dot{a}(\rho+ p)=0.
\end{align}

As usual, $N(t)$ is not determined by the dynamical equations. Instead, this function must be specified by a choice of gauge, which fixes the physical meaning of the coordinate $t$. An intuitive representation of the dynamical equations is achieved by setting $N(t)=1/a(t)$, so that $t$ is the proper time $\tau$ of a comoving observer in the FRW spacetime (\ref{FRWmetric}). The semiclassical Friedmann equations (\ref{FEgen}) then become
\begin{subequations}\label{FEH}
\begin{align}\label{FEH1}
H^2 &= \frac{\kappa}{3}   \rho 
 - \left(1 +  \frac{4 \bar{g}}{15 } \right) \frac{k} {a^2}
+   \frac{\bar{g} a^{-2}}{\left(\int^{\bar{\tau}}_{\tau} \frac{\ud \tau'}{a(\tau')}\right)^{2}}
- \frac{2\bar{g}}{a^4}\int^{\tau}_{\tau_-} \ud \tau''    \frac{a(\tau'')}{\left(\int^{\bar{\tau}}_{\tau''}\frac{\ud \tau'}{a(\tau')}\right)^{3}},
\\\label{FEH2}
\frac{\ud H}{\ud \tau } + 2 H^2&=\frac{\kappa}{6}  \left(\rho-3p \right)
 - \left(1 +  \frac{4 \bar{g}}{15 } \right)\frac{k}{a^2}   
 +   \frac{\bar{g}a^{-2}}{\left(\int^{\bar{\tau}}_{\tau} \frac{\ud \tau'}{a(\tau')}\right)^{2}},
\end{align}
\end{subequations}
where $H\equiv \ud \ln a/\ud \tau$ is the Hubble parameter. Subtracting (\ref{FEH1}) from (\ref{FEH2}) we can also obtain the acceleration equation:
\begin{align}\label{accn}
\frac{1}{a}\frac{\ud^2 a }{\ud \tau^2}= \frac{\ud H}{\ud \tau } + H^2 =-\frac{\kappa}{6}   \left(\rho+3p \right)+\frac{2\bar{g}}{a^4}\int^{\tau}_{\tau_-} \ud \tau''    \frac{a(\tau'')}{\left(\int^{\bar{\tau}}_{\tau''}\frac{\ud \tau'}{a(\tau')}\right)^{3}}.
\end{align}
This confirms our basic hypothesis -- quantum bias (\ref{Vholo}) does indeed generate cosmic acceleration, without the need for a cosmological constant, dark energy, or modified gravity. Note that $\bar{g} >0$ gives quantum bias the correct sign, producing \emph{positive} cosmic acceleration. This sign is guaranteed by the symmetry-breaking argument of appendix \ref{Gauge}: we are forced to set $\xi=1/4$ in the definition (\ref{dbardef}) and hence restrict $\bar{d}\in \mathbb{N}$ for the minimal model (\ref{DoFmetric}) or $\bar{d}\in \mathbb{R}^+$ for the generalisation (\ref{DoFmetricShape}); in either case, we have $\bar{g}\equiv \pi^2 \bar{\mu}^2/\bar{d}>0$.

To study this new form of cosmic acceleration (\ref{accn}) in detail, we must of course solve the semiclassical Friedmann equations. To this end, the gauge $N(t)=1$ is an extremely profitable choice: $t$ is then equivalent to conformal time (\ref{etadef}) and the semiclassical Friedmann equations (\ref{FEgen}) simplify to
\begin{subequations}\label{FEeta}
\begin{align}\label{FEeta1}
\left(\frac{\ud a}{\ud \eta} \right)^2 &= \frac{\kappa}{3}  \rho  a^4
 - \left(1 +  \frac{4\bar{g}}{15} \right) k  a^2
+   \frac{\bar{g} a^2}{\left(\bar{\eta}-\eta \right)^{2}}
- 2\bar{g}\int^{\eta}_{\eta_-} \ud \eta'    \frac{ [a(\eta')]^2}{\left(\bar{\eta}-\eta'\right)^{3}},
\\\label{FEeta2}
\frac{\ud^2 a}{\ud \eta^2} &=\frac{\kappa}{6}  \left(\rho-3p \right)a^3 
 - \left(1 +  \frac{4\bar{g}}{15} \right)k a  
 + \frac{\bar{g} a}{\left(\bar{\eta}-\eta\right)^{2}} .
\end{align}
\end{subequations}
In the next section, we will find exact solutions to these equations, for $k=0$.
\newpage
\end{widetext}

\section{Spatially Flat Universe with\\ Matter \& Radiation}\label{Solutions}
Let us model the universe as a spatially flat FRW spacetime (\ref{FRWmetric}) containing   pressure-free matter (so-called ``dust'') and radiation. In other words, we set $k=0$ and
\begin{align}
\rho &= \frac{\rho_\mathrm{m0} a_{0}^3}{a^3} + \frac{\rho_\mathrm{r0} a_{0}^4}{a^4},&
p&= \frac{\rho_\mathrm{r0} a_{0}^4}{3 a^4}.
\end{align}
Here, $\rho_\mathrm{m0}$ is the energy-density of matter, and $\rho_\mathrm{r0}$  the energy-density of radiation, when the scale factor has some arbitrary reference value $a=a_{0}$. (Typically, we interpret $\{a_0, \rho_\mathrm{m0}, \rho_\mathrm{r0}\}$ as ``present-day'' values.) The semiclassical Friedmann equations (\ref{FEeta}) are therefore 
\begin{subequations}\label{FEetaflat}
\begin{align}\nonumber
\left(\frac{\ud a}{\ud \eta} \right)^2 &= \frac{\kappa }{3}\left(\rho_\mathrm{m0} a_{0}^3a +\rho_\mathrm{r0}a_{0}^4\right) 
+   \frac{\bar{g} a^2}{\left(\bar{\eta}-\eta \right)^{2}}\\\label{FEetaflat1}
&\quad {}- 2\bar{g}\int^{\eta}_{0} \ud \eta'    \frac{ [a(\eta')]^2}{\left(\bar{\eta}-\eta'\right)^{3}},
\\\label{FEetaflat2}
\frac{\ud^2 a}{\ud \eta^2} &=\frac{\kappa}{6} \rho_\mathrm{m0} a_{0}^3
 + \frac{\bar{g} a}{\left(\bar{\eta}-\eta\right)^{2}},
\end{align}
\end{subequations}
where the cutoff $\eta_-$ has been placed at the Big Bang:
\begin{align}\label{BB}
\eta_-&=0, & \lim_{\eta \to 0}a(\eta)&=0.
\end{align}
As with our preceding analysis, we ignore the details of the very early universe, including inflation and the possibility of a Big Bounce.\footnote{The  behaviour of $a(\eta)$ at very early times (e.g.\ during inflation) will slightly affect the value of the integral in equation (\ref{FEetaflat1}); however, this section of the integral is far smaller than all the other terms, and can safely be neglected. (We will prove this in a future publication, when we cover the very early universe in detail.) As such, the post-inflationary universe (\ref{FEetaflat}) can be treated as though it began with a classical Big Bang (\ref{BB}).}

\subsection{Derivation}
Let us first simplify our notation. We define the constants
\begin{align}\label{betadefs}
 \beta_\mathrm{m}&\equiv \frac{\bar{\eta}^2 \kappa  \rho_\mathrm{m0} a^3_{0}}{3},& \beta_\mathrm{r}&\equiv \frac{ \bar{\eta}^2\kappa  \rho_\mathrm{r0} a^4_{0}}{3},
\end{align}
and express the conformal time in terms of the variable
\begin{align}\label{udef}
u \equiv \frac{\bar{\eta}-\eta}{\bar{\eta}}.
\end{align}
This recasts the dynamical equations (\ref{FEetaflat}) as
\begin{subequations}\label{FEsimple}
\begin{align}\label{FEsimple1}
\left(\frac{\ud a}{\ud u} \right)^2  &= \beta_\mathrm{m}a   +  \beta_\mathrm{r} 
 +   \frac{\bar{g} a^2}{u^2}
 - 2\bar{g}\int^{1}_{u} \ud u'    \frac{ [a(u')]^2}{u'^3},
\\\label{FEsimple2}
\frac{\ud^2 a}{\ud u^2} &=\frac{\beta_\mathrm{m}}{2}  + \frac{\bar{g}a}{u^{2}},
\end{align}
\end{subequations}
which we shall now proceed to solve.

To obtain the general solution of (\ref{FEsimple2}), note that the homogeneous equation 
\begin{align}
\frac{\ud^2 a}{\ud u^2} = \frac{\bar{g} a}{u^{2}}
\end{align}
has general solution
\begin{align}
a = C_+ u^{\left(1 +\sqrt{4\bar{g}+1}\right)/2} + C_- u^{\left(1 -\sqrt{4\bar{g}+1}\right)/2},
\end{align}
for arbitrary constants $C_\pm$. Let us write this as
\begin{align}\label{homogensol}
a = C_+ u^{\left(1 +\bar{\gamma} \right)/2} + C_- u^{\left(1 -\bar{\gamma}\right)/2},
\end{align}
where
\begin{align}\label{gammadef}
\bar{\gamma}&\equiv\sqrt{4 \bar{g} +1}=\sqrt{\frac{4 \pi^2 \bar{\mu}^2}{\bar{d}} +1}
\end{align}
repackages the unknown constant $\bar{d}$ in a convenient fashion. We will generally be interested in $\bar{\gamma}>1$, which corresponds to positive cosmic acceleration: $\bar{g}>0$ in equation (\ref{accn}). Beyond this, the solutions (\ref{homogensol}) remain well-defined for all $\bar{g}\ge-1/4$, and we can take $\bar{\gamma}\ge 0$ without loss of generality. (As there are no real solutions for $\bar{g}<-1/4$, such values are completely untenable.) 

In addition to the homogeneous solutions (\ref{homogensol}) we require a particular integral. It is easy to check that
\begin{align}
a= \frac{\beta_\mathrm{m}}{4 -2\bar{g}}u^2=\frac{2\beta_\mathrm{m}}{9-\bar{\gamma}^2}u^2
\end{align}
satisfies the second semiclassical equation (\ref{FEsimple2}); hence
\begin{align}\label{gensol}
a= \frac{2\beta_\mathrm{m}}{9-\bar{\gamma}^2}u^2 + C_+ u^{\left(1 +\bar{\gamma} \right)/2} + C_- u^{\left(1 -\bar{\gamma}\right)/2}
\end{align}
is its general solution.

We now impose the following conditions on the scale factor (\ref{gensol}):
\begin{subequations}\label{acond}
\begin{align}\label{acond1}
\left. a\right|_{u=1}&=0,\\\label{acond2}
\left.\frac{\ud a}{\ud u}\right|_{u=1}&= -\sqrt{\beta_\mathrm{r}}.
\end{align}
\end{subequations}
The first equation (\ref{acond1}) is simply the Big Bang condition (\ref{BB}) expressed in terms of $u$. The second (\ref{acond2}) ensures that the other Friedmann equation (\ref{FEsimple1}) is satisfied at $u=1$, with the negative root providing an expanding universe: $\ud a/\ud \eta >0$. In fact, this condition guarantees that (\ref{FEsimple1}) is satisfied for all $u$. To see this clearly, move all the terms in (\ref{FEsimple1}) to one side of the equation, and call this sum $E(u)$. Differentiating with respect to $u$, one finds that $E'(u)$ vanishes whenever (\ref{FEsimple2}) is satisfied, so our solution (\ref{gensol}) guarantees $E'(u)=0\ \forall u$. Given that  (\ref{acond2}) sets $E(1)=0$, we conclude that $E(u)=E(1)- \int_u^1E'(u')\ud u'= 0$, meaning that equation (\ref{FEsimple1}) is satisfied for all $u$. Thus, the conditions (\ref{acond}) ensure that our solution (\ref{gensol}) solves \emph{both} semiclassical Friedmann equations (\ref{FEsimple}) and has a Big Bang at $\eta=0$.

Inserting (\ref{gensol}) into (\ref{acond}) we obtain
\begin{subequations}
\begin{align}
\frac{2\beta_\mathrm{m}}{9-\bar{\gamma}^2}+ C_+ + C_- &=0,\\
\frac{4 \beta_\mathrm{m}}{9-\bar{\gamma}^2}+\frac{1+\bar{\gamma}}{2}  C_+ +\frac{1-\bar{\gamma}}{2} C_- &= -\sqrt{\beta_\mathrm{r}},
\end{align}
\end{subequations}
and hence
\begin{align}\label{coefs}
C_\pm = \mp \frac{1}{\bar{\gamma}}\left( \frac{\beta_\mathrm{m}}{3\mp \bar{\gamma}} + \sqrt{\beta_\mathrm{r}}\right).
\end{align}
Substituting these coefficients back into equation (\ref{gensol}) we obtain the general solution:
\begin{align}\nonumber
a&= \frac{\beta_\mathrm{m}}{\bar{\gamma}}\left( \frac{2 \bar{\gamma} u^2}{9-\bar{\gamma}^2} - \frac{u^{\left(1 +\bar{\gamma} \right)/2}}{3-\bar{\gamma}} + \frac{u^{\left(1 -\bar{\gamma}\right)/2}}{3+\bar{\gamma}}\right) \\\label{asol}
&\quad {} - \frac{\sqrt{\beta_\mathrm{r}}}{\bar{\gamma}}\left(u^{\left(1 +\bar{\gamma}\right)/2}- u^{\left(1 -\bar{\gamma}\right)/2} \right),
\end{align}
which also determines the proper time  since the Big Bang:
\begin{align}\nonumber
\tau &= \int_{0}^\eta  \ud \eta' a(\eta') = \bar{\eta}\int^{1}_u  \ud u' a(u')\\ \nonumber
&= \frac{2\bar{\eta} \beta_\mathrm{m}}{\bar{\gamma}\left(9-\bar{\gamma}^2\right)}\left( \frac{ \bar{\gamma}}{3} \left(1-u^3\right)+ u^{\left(3 +\bar{\gamma}\right)/2} - u^{\left(3 -\bar{\gamma}\right)/2}\right)\\\label{tausol}
&\quad {} +  \frac{2 \bar{\eta}\sqrt{\beta_\mathrm{r}}}{\bar{\gamma}}\left( \frac{2\bar{\gamma}}{9-\bar{\gamma}^2} +\frac{u^{\left(3 +\bar{\gamma}\right)/2}}{3+\bar{\gamma}} -\frac{u^{\left(3 -\bar{\gamma}\right)/2}}{3-\bar{\gamma}} \right).
\end{align}
This completes the task of solving the semiclassical Friedmann equations (\ref{FEgen}). Equations (\ref{asol}) and (\ref{tausol}) are parametric solutions $a=a(u)$, $\tau=\tau(u)$, $u\in[0,1]$, that generate the expansion history $a(\tau)$ of a spatially flat universe (containing matter and radiation) accelerated by holographic quantum bias (\ref{Vholo}). In addition to $\{a(u),\tau(u)\}$ we can also write down a simple parametric expression for the conformal time that has elapsed since the Big Bang:
\begin{align}\label{etasol}
\eta = \bar{\eta}\, (1-u),
\end{align}
as follows directly from the definition of $u$ (\ref{udef}). In the next section, we will express results (\ref{asol}--\ref{etasol}) in a more useful form, and extract the behaviour of key cosmological observables.

\subsection{Cosmological Solutions}
For the sake of brevity, we write the parametric solutions (\ref{asol}--\ref{etasol}) as
\begin{subequations}\label{fullsol}
\begin{align}\label{fullsola}
&&a&=\beta_\mathrm{m}\left[\FG{\prime}{u}\right],\\\label{fullsoltau}
u& \in [0,1]:& \tau&=-\bar{\eta}\beta_\mathrm{m}\left[\FG{}{u}\right],\\\label{fullsoleta}
&& \eta &= \bar{\eta}\, (1-u),
\end{align}
\end{subequations}
having introduced the functions
\begin{align}\nonumber
F_{\bar{\gamma}}(u)&\equiv \frac{-2}{\bar{\gamma}\left(9-\bar{\gamma}^2\right)}\left( \frac{ \bar{\gamma}}{3} \left(1-u^3\right)+ u^{\left(3 +\bar{\gamma}\right)/2} - u^{\left(3 -\bar{\gamma}\right)/2}\right) ,\\ \label{FGdef}
G_{\bar{\gamma}}(u)&\equiv \frac{-2}{\bar{\gamma}}\left( \frac{2\bar{\gamma}}{9-\bar{\gamma}^2}+\frac{u^{\left(3 +\bar{\gamma}\right)/2}}{3+\bar{\gamma}} -\frac{u^{\left(3 -\bar{\gamma}\right)/2}}{3-\bar{\gamma}} \right),
\end{align}
and the ratio
\begin{align}\label{alphadef}
\alpha\equiv\sqrt{\beta_\mathrm{r}}/\beta_\mathrm{m}.
\end{align}
The aim of this section is to eliminate the unfamiliar quantities $\{\bar{\eta},\beta_\mathrm{m},\alpha\}$ and connect the solutions (\ref{fullsol}) to standard cosmological observables.

Consulting definitions (\ref{betadefs}) and (\ref{alphadef}), we begin by expressing the density parameters as follows:
\begin{align}\bs\label{OmegaCalc}
\Omega_\mathrm{m}& \equiv \frac{\kappa \rho_\mathrm{m}}{3H^2}= \frac{\kappa \rho_\mathrm{m0}a_0^3/a^{3}}{3H^2}= \frac{\beta_\mathrm{m}}{ a}\cdot\frac{1}{\left(a\bar{\eta} H\right)^2},\\
\Omega_\mathrm{r}&\equiv \frac{\kappa \rho_\mathrm{r}}{3H^2} = \frac{\kappa \rho_\mathrm{r0}a_0^4/a^{4}}{3H^2}=\left( \frac{\alpha \beta_\mathrm{m}}{ a}\right)^2\cdot\frac{1}{\left(a\bar{\eta} H\right)^2}.\es
\end{align}
Notice that the factors on the right can be calculated directly from the expansion histories (\ref{fullsol}): clearly $\beta_\mathrm{m}/a=[\FG{\prime}{u}]^{-1}$, and 
\begin{align}
a \bar{\eta} H = \bar{\eta}\,\frac{\ud a}{\ud \tau} =- \frac{\FG{\prime\prime}{u}}{\FG{\prime}{u}}.
\end{align}
Hence, the densities (\ref{OmegaCalc}) become
\begin{subequations}\label{Omega}
\begin{align}\label{Omegam}
\Omega_\mathrm{m}&=\frac{\FG{\prime}{u}}{\left[\FG{\prime\prime}{u}\right]^2},\\\label{Omegar}
 \Omega_\mathrm{r}&=\frac{\alpha^2}{\left[\FG{\prime\prime}{u}\right]^2}.
\end{align}
\end{subequations}
When evaluated at the current time $u=u_0$, the above formulae determine the present-day density parameters $\{\Omega_{\mathrm{m}0},\Omega_{\mathrm{r}0}\}$. As such, equations (\ref{Omega}) allow us to convert the new variables $\{u_0,\alpha\}$ into standard observables $\{\Omega_{\mathrm{m}0},\Omega_{\mathrm{r}0}\}$ for each value of the fundamental constant $\bar{\gamma}$. In fact, we can solve equation (\ref{Omegar}) explicitly:\footnote{To ensure the correct sign when taking the square root of equation (\ref{Omegar}), consider the Big Bang limit $u\to 1$, where $\Omega_\mathrm{r}\to 1$, $F^{\prime\prime}_{\bar{\gamma}}(u)\to 0$ and $G^{\prime\prime}_{\bar{\gamma}}(u)\to -1$.}
\begin{align}\label{alphasolve}
\alpha&= \frac{-F^{\prime\prime}_{\bar{\gamma}}(u_0)}{\left(\Omega_\mathrm{r0}\right)^{-1/2}+ G^{\prime\prime}_{\bar{\gamma}}(u_0)},
\end{align}
which allows us to eliminate $\alpha$ whenever we wish. Inserting this result into equation (\ref{Omegam}) we then obtain a formula $\Omega_{\mathrm{m0}}=\Omega_\mathrm{m0}(u_0,\Omega_\mathrm{r0},\bar{\gamma})$, which implicitly  relates $u_0$ to $\{\Omega_\mathrm{m0}, \Omega_\mathrm{r0}, \bar{\gamma}\}$. However, without a closed-form solution $u_0=u_0(\Omega_\mathrm{m0},\Omega_\mathrm{r0},\bar{\gamma})$ we cannot completely eliminate $u_0$ from our formalism. Instead, it is convenient to keep $u_0$ as a basic cosmological parameter -- fixing  the observer's ``present day'' -- and determine $\Omega_\mathrm{m0}$ with equation (\ref{Omegam}). 

With this in mind, we return to the solutions (\ref{fullsol}) and study their behaviour at $u=u_0$. In particular, we see 
\begin{align}\label{betamtoa0}
a_0 &= \beta_\mathrm{m}\left[\FG{\prime}{u_0}\right],\\ 
H_0 &= \left(\frac{1}{a}\frac{\ud a}{\ud \tau}\right)_{u_0}= \frac{-\left[\FG{\prime\prime}{u_0}\right]}{\bar{\eta}\beta_\mathrm{m}\left[\FG{\prime}{u_0}\right]^2}.
\end{align}
Solving these relations for $\bar{\eta}$ and $\beta_\mathrm{m}$, and substituting the result back into the solutions (\ref{fullsol}), we arrive at a particularly useful representation of the predicted expansion histories:
\begin{subequations}\label{exphist}
\begin{align}\label{exphista}
\frac{a}{a_0}&=\frac{\FG{\prime}{u}}{\FG{\prime}{u_0}},\\\label{exphisttau}
\tau&= \frac{\left[\FG{}{u}\right]\left[\FG{\prime\prime}{u_0}\right]}{H_0\left[\FG{\prime}{u_0}\right]^2},\\\label{exphisteta}
a_0 \eta  &= \frac{u-1}{H_0}\cdot \frac{\FG{\prime\prime}{u_0}}{\FG{\prime}{u_0}},
\end{align}
with the Hubble parameter given by
\begin{align}\label{exphistH}
\frac{H}{H_0}&= \frac{\FG{\prime\prime}{u}}{\FG{\prime\prime}{u_0}}\!\left[\frac{\FG{\prime}{u_0}}{\FG{\prime}{u}}\right]^2\!.\!
\end{align}
\end{subequations}
Equations (\ref{Omega}) and (\ref{exphist}) represent the main predictions of the theory, applicable to a spatially flat universe containing matter and radiation. For given values $\{H_0,u_0,\Omega_\mathrm{r0},\bar{\gamma}\}$, these results describe the evolution of the scale factor,  proper time,  conformal time, and  matter/radiation densities, as a function of $u$: from the Big Bang $u=1$, to present day $u=u_0$, and into the distant future $u\to 0$. Recall that $F_{\bar{\gamma}}$ and $G_{\bar{\gamma}}$ are defined in (\ref{FGdef}), $\alpha$ is set by equation (\ref{alphasolve}), and 
\begin{align}
\bar{\gamma}&\equiv\sqrt{4\bar{g} +1}= \sqrt{\frac{4\pi^2 \bar{\mu}^2}{\bar{d}} +1}
\end{align}
is a fundamental constant. (The numerical factor $\bar{\mu}=1/(24 \ln 2 -15) \approx 0.61142$ accounts for the arrangement of holographic information in spacetime -- see appendix \ref{HoloPrinc}. The parameter $\bar{d}$ depends on unknown details of the discarded configuration space (\ref{dbardef}) but may be constrained to $\bar{d}>0$, or even $\bar{d} \in \mathbb{N}$, by the invariance argument of Appendix \ref{Gauge}.) For each expansion history (\ref{exphist}) in the theoretically well-motivated class $\bar{\gamma}>1$, the universe undergoes positive late-time acceleration (\ref{accn}) due to the quantum bias (\ref{Vholo}) from its holographic information capacity (\ref{Sholo}).

For the remainder of this section, we will study the basic properties of the predicted cosmologies (\ref{exphist}); then, in section \ref{Compare}, we will make a detailed comparison with the expansion histories of the standard $\Lambda$CDM model.

\subsection{Limiting Values of $\bar{\gamma}$}
At first glance, the functions (\ref{FGdef}) appear to break down at $\bar{\gamma}=0$ and $\bar{\gamma} =3$. In fact, the limits $\bar{\gamma}\to 0$ and $\bar{\gamma}\to 3$ are entirely well-behaved:
\begin{subequations}\label{limitsol}
\begin{align}\bs\label{limitsol0}
\lim_{\bar{\gamma}\to 0}F_{\bar{\gamma}}(u)&=\frac{-2}{27}\left(1- u^3 +3 u^{3/2}\ln u\right),\\
\lim_{\bar{\gamma}\to 0}G_{\bar{\gamma}}(u)&= \frac{-2}{9}\left(2+ u^{3/2}\left(3\ln u -2 \right) \right),\es
\\
\bs\label{limitsol3}
\lim_{\bar{\gamma}\to 3}F_{\bar{\gamma}}(u)&=\frac{1}{54}\left(2\left(1-u^3\right)+3\left(1+u^3\right)\ln u\right), \\
\lim_{\bar{\gamma}\to 3}G_{\bar{\gamma}}(u)&= \frac{1}{9}\left(1-u^3 + 3 \ln u\right).\es
\end{align}
\end{subequations}
Hence, the expansion histories (\ref{exphist}) exist for all $\bar{\gamma} \ge 0$.

\subsection{Final Conformal Time}\label{etabarcheck}
We are now in a position to check the self-consistency of the theory, confirming that $\bar{\eta}$ really is the \emph{final} conformal time (\ref{etabardef}). Evaluating our solutions (\ref{fullsol}) in the limit $u\to 0$, we see that
\begin{align}\label{finaltimes}\bs
\lim_{\eta\to \bar{\eta}}a&=  \begin{cases}
0, & \bar{\gamma}\in[0,1), \\
\text{finite}, &  \bar{\gamma}= 1, \\
\infty, &  \bar{\gamma}\in (1,\infty),\\
\end{cases}\\
\lim_{\eta\to \bar{\eta}}\tau &=  \begin{cases}
\text{finite}, &   \bar{\gamma}\in[0,3), \\
\infty , &   \bar{\gamma}\in[3,\infty).
\end{cases}\es
\end{align}
For the well-motivated values $\bar{\gamma}>1$, we recover exactly what we need: an accelerating expanding universe that attains infinite expansion as $\eta$ approaches $\bar{\eta}$. For $\bar{\gamma} \in (1,3)$ the universe ends in a  Big Rip in finite proper time, while for $\bar{\gamma} \in [3,\infty)$ the limit $\eta\to \bar{\eta}$ is achieved asymptotically as $\tau \to \infty$. In the next subsection, we will interpret these behaviours in terms of an effective equation of state $w_\mathrm{eff}$ for holographic quantum bias.

Before then, let us quickly comment on the remaining (unphysical) values $\bar{\gamma} \in [0,1]$. For $\bar{\gamma} \in [0,1)$ the universe ends in a Big Crunch at $\eta=\bar{\eta}$. These solutions pass the basic consistency check ($\bar{\eta}$ is indeed the final conformal time) but violate the assumption of an expanding universe $\dot{a}>0$. This assumption was used to derive the information capacity (\ref{Sholo}) so the physical self-consistency of these solutions remains dubious. Finally, there is the trivial value $\bar{\gamma}=1$, which sets $\bar{g}=0$ and reduces the semiclassical Friedmann equations (\ref{FEetaflat}) to the \emph{classical} Friedmann equations. These formulae make no reference to $\bar{\eta}$, so nothing special happens at $\eta=\bar{\eta}$ in this case.

\subsection{Effective Equation of State}\label{EoS}
It is often useful to think of quantum bias as though it were a homogeneous fluid, contributing an effective energy-density $\rho_\mathrm{eff}$ and pressure $p_\mathrm{eff}$ to the classical Friedmann equations. Consulting the semiclassical Friedmann equations (\ref{FEeta}) for $k=0$, we see that this fictitious fluid must have
\begin{align}\bs \label{rhoandp}
\kappa \rho_\mathrm{eff}&=  \frac{3\bar{g} }{a^2\left(\bar{\eta}-\eta \right)^{2}}
- \frac{6\bar{g}}{a^4}\int^{\eta}_{0} \ud \eta'    \frac{ [a(\eta')]^2}{\left(\bar{\eta}-\eta'\right)^{3}},\\
\kappa p_\mathrm{eff}&=  - \frac{\bar{g} }{a^2\left(\bar{\eta}-\eta \right)^{2}}
- \frac{2\bar{g}}{a^4}\int^{\eta}_{0} \ud \eta'    \frac{ [a(\eta')]^2}{\left(\bar{\eta}-\eta'\right)^{3}},\es
\end{align}
and equation of state
\begin{align}\label{wdef}
w_\mathrm{eff}\equiv \frac{p_\mathrm{eff}}{\rho_\mathrm{eff}}&= -\frac{\frac{a^2}{(\bar{\eta}-\eta)^2} + 2 \int^{\eta}_{0} \ud \eta'    \frac{ [a(\eta')]^2}{\left(\bar{\eta}-\eta'\right)^{3}} }{3\left(\frac{a^2}{(\bar{\eta}-\eta)^2}- 2\int^{\eta}_{0} \ud \eta'    \frac{ [a(\eta')]^2}{\left(\bar{\eta}-\eta'\right)^{3}} \right)}.
\end{align}
However, this description should not be taken too literally: there is nothing to suggest that $\{\rho_\mathrm{eff},p_\mathrm{eff}\}$ can be interpreted \emph{locally} in terms of a physical fluid. Indeed, the cosmological quantum bias (\ref{Vholo}) only applies to a volume $\mathscr{V}_*$ much larger than the cosmological event horizon, so there is little reason to believe in variations $\{\delta\rho_\mathrm{eff},\delta p_\mathrm{eff}\}$ below this length-scale. As such, we should treat the effective dark fluid as a purely \emph{global} phenomenon, which only affects the behaviour of matter perturbations via the evolution of the background $a(\tau)$.

To apply this formalism to our exact solutions (\ref{exphist}) we first rewrite the equation of state (\ref{wdef}) in terms of the variable $u$:
\begin{align}\label{weffu}
w_\mathrm{eff}&=-\frac{ a^2u^{-2} +2\int^{1}_{u} \ud u' [a(u')]^2u'^{-3} }{3\left(a^2u^{-2}- 2 \int^{1}_{u} \ud u'    [a(u')]^2u'^{-3}\right) }.
\end{align}
At early times, we can write $u=1-\epsilon$ and expand the scale factor (\ref{exphista}) in powers of $\epsilon \equiv \eta/\bar{\eta}$; using $F'_{\bar{\gamma}}(1)=G'_{\bar{\gamma}}(1)=F''_{\bar{\gamma}}(1)=G'''_{\bar{\gamma}}(1)=0$, $G''_{\bar{\gamma}}(1)=-1$ and $F'''_{\bar{\gamma}}(1)=1/2$, we obtain
\begin{align}\label{aexpand}
\frac{a}{a_0}&=\frac{ \alpha \epsilon + \epsilon^2/4  + O\!\left(\epsilon^3\right)}{\FG{\prime}{u_0}}.
\end{align}
Substituting this expansion into the equation of state (\ref{weffu}) we find
\begin{align}\label{wearly}
 w_\mathrm{eff}&=-\frac{1}{3}- \frac{4}{9}\epsilon+O(\epsilon^2).
\end{align}
In other words, quantum bias behaves like spatial curvature $w_k=-1/3$, as we approach the initial singularity. Intuitively, this is because the integrals in equation (\ref{rhoandp}) are small compared to the terms proportional to $1/(\bar{\eta}-\eta)^{2}a^{2} \approx 1/\bar{\eta}^{2}a^{2}$.

At late times, however, the integrals cannot be neglected. Considering $\eta \to \bar{\eta}$, $u\to0$, the solutions (\ref{exphista}) behave as follows:
\begin{align}
a \propto u^{\left(1-\bar{\gamma}\right)/2}\left(1 + O\left(u^{\min \left\{\bar{\gamma},\left(3+\bar{\gamma}\right)/2\right\}}\right)\right),
\end{align}
for $\bar{\gamma}>1$. Hence, the equation of state (\ref{weffu}) tends to
\begin{align}\label{wlate}
\lim_{\eta \to \bar{\eta}} w_\mathrm{eff}&=\frac{3 + \bar{\gamma}}{3(1-\bar{\gamma})}.
\end{align}
For $\bar{\gamma} \in (1,3)$, we see that quantum bias resembles \emph{phantom} dark energy ($w_\mathrm{eff}<-1$) at late times, explaining the Big Rips in equation (\ref{finaltimes}). For these solutions (\ref{exphist}) the physical area of the cosmological event horizon $A_{\mathrm{EH}}=4\pi [a(\eta)]^2(\bar{\eta}-\eta)^2\sim u^{3-\bar{\gamma}}\to 0$ at late times ($u\to 0$) causing $\rho_\mathrm{eff}\sim 1/A_\mathrm{EH}$ to grow without bound. The other values $\bar{\gamma} \in (3,\infty)$ generate non-phantom  behaviour ($-1<w_\mathrm{eff}<-1/3$) at late times, which accelerates the universe over unbounded proper time. We also note that the special case $\bar{\gamma}=3$ has $w_\mathrm{eff}\to-1$, converging on the equation of state of a cosmological constant. Hence the special solution (\ref{limitsol3}) must tend to de Sitter spacetime in the asymptotic future.

\begin{figure}
    \centering
\includegraphics[width=0.49\textwidth]{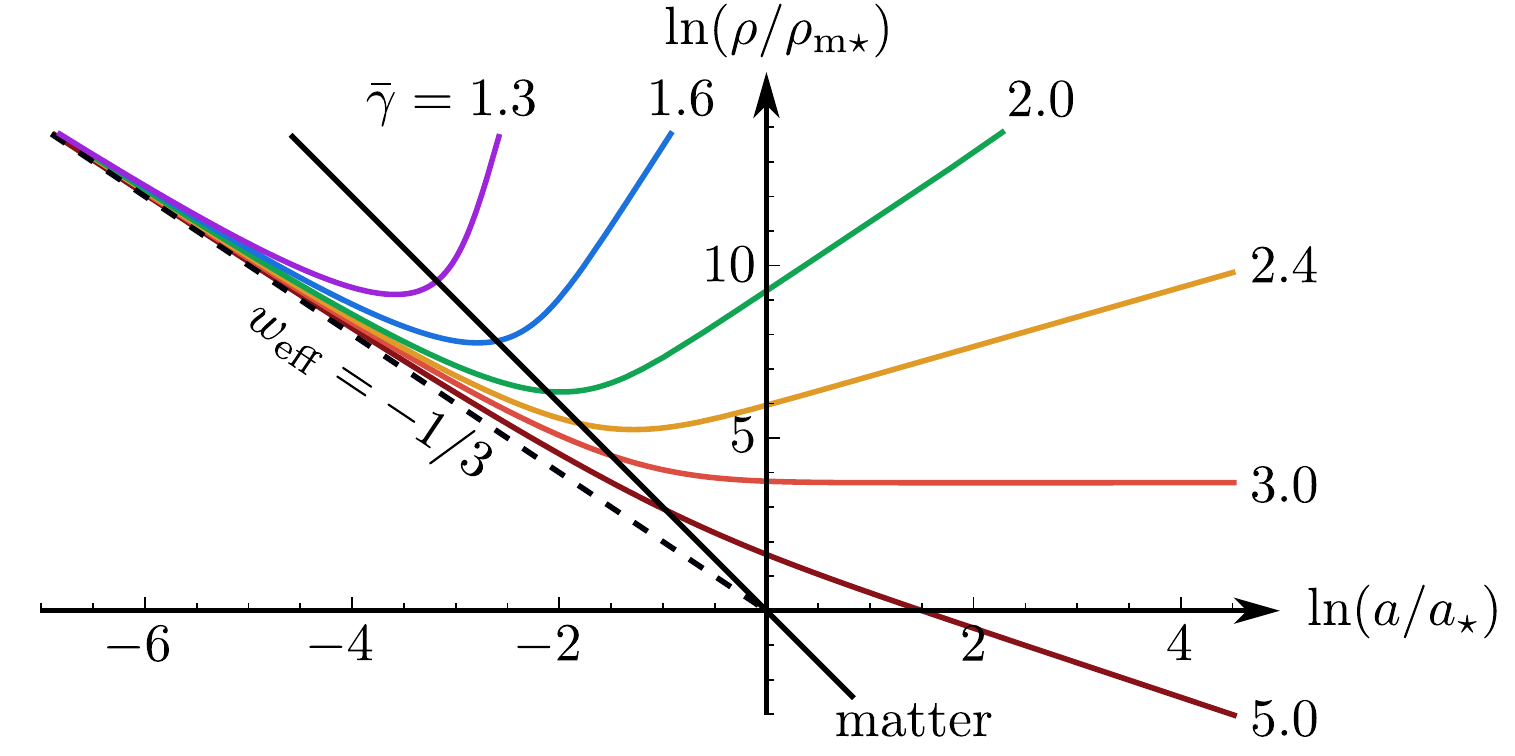}
\vspace{-0.3cm}
\caption{The above graph depicts the behaviour of the matter density $\rho_\mathrm{m}$ (black line) and the effective dark energy density $\rho_\mathrm{eff}$ (coloured lines) as the universe expands. For the sake of clarity, we neglect radiation ($\Omega_\mathrm{r}=0\, \Leftrightarrow\, \alpha=0$) and use reference values $\{\rho_\mathrm{m\star},a_\star\}$ such that the early-time asymptote passes through the origin, for each value of $\bar{\gamma}$. (Specifically: $\rho_\mathrm{m\star}=\rho_\mathrm{m}(u_\star)$, $a_\star=a(u_\star)$, with $u_\star$ solving $F^\prime_{\bar{\gamma}}(u_\star)=4/(\bar{\gamma}^2-1)$.) In general, the early-time behaviour (\ref{wearly}) is accurate during the matter-dominated era $\rho_\mathrm{eff}\ll\rho_\mathrm{m}$, but breaks down as $\rho_\mathrm{eff}$ approaches $\rho_\mathrm{m}$. While $\rho_\mathrm{eff}\approx\rho_\mathrm{m}$, the effective equation of state $w_\mathrm{eff}$ becomes more negative, and hence the gradient $\ud \ln \rho_\mathrm{eff}/\ud \ln a$ increases. Ultimately, quantum bias dominates $\rho_\mathrm{eff}\gg\rho_\mathrm{m}$, and $w_\mathrm{eff}$ converges on its final value (\ref{wlate}). For $\bar{\gamma}\in (1,3)$, the dark energy density $\rho_\mathrm{eff}$ always has a turning point when $\rho_\mathrm{eff}\approx \rho_\mathrm{m}$; the neighbourhood of this minimum resembles the current state of our universe: $\Omega_\mathrm{m}\approx 1/2$, $w_\mathrm{eff}\approx-1$. }\label{DensityPlot}
\end{figure}

The transition from early times (\ref{wearly}) to late times (\ref{wlate}) is illustrated in figure \ref{DensityPlot}. For numerical calculations, it is  often useful to  eliminate the integrals from formula (\ref{weffu}) using the first semiclassical Friedmann equation (\ref{FEsimple1}). If we then insert the scale factor solution (\ref{fullsola}) we arrive at
\begin{align}\label{weffsol}
w_\mathrm{eff}&= \frac{\left(\bar{\gamma}^2-1\right)\left[\FGn{\prime}\right]^2}{6 u^2\left(\alpha^2 +\FGn{\prime}- \left[\FGn{\prime\prime}\right]^2\right)}+\frac{1}{3},\!
\end{align}
which is a purely algebraic function of $\{u,\alpha,\bar{\gamma}\}$.

\section{Comparison with $\Lambda$CDM}\label{Compare}
Rather than attempt a full comparison with observational data here, we can assess the plausibility of the theory by comparing its predicted expansion histories (\ref{exphist}) to those of $\Lambda$CDM. This should assuage fears that the model can be dismissed ``out of hand'' as inconsistent with observations. 

A few notes before we start our comparison:
\begin{itemize}
\item We will ignore radiation ($\Omega_\mathrm{r}=0$) in the following analysis. This approximation is sufficient to describe the universe as far back as recombination $a=a_*\approx a_0/1100 $, when quantum bias will be seen to be negligble: $\rho_\mathrm{eff*}/\rho_\mathrm{m*}<10^{-4}$. We can also be sure that $\rho_\mathrm{eff}$ is irrelevant at earlier times, due to its primordial equation of state (\ref{wearly}). 

\item \emph{Notation}: We shall refer to the new theory as  Quantum Bias Cosmology, or QB Cosmology. Here we will study QB-CDM cosmologies, which include the standard cold dark matter component. I shall distinguish $\Lambda$CDM quantities from QB-CDM quantities with superscripts $(\Lambda)$ and (QB).
\end{itemize}
We now begin by describing the behaviour of the standard $\Lambda$CDM universe.

\subsection{$\Lambda$CDM Cosmology}
According to the \emph{classical} Friedmann equations, a flat universe $k=0$, containing only matter $\rho_m\propto a^{-3}$ and a cosmological constant $\Lambda>0$, expands according to 
\begin{subequations}\label{LCDM}
\begin{align}\label{aLambda}
\frac{a^{(\Lambda)}}{a^{(\Lambda)}_0}&= \left[\frac{\sinh (v)}{\sinh (v_0)}\right]^{2/3},
\\ \label{tauLambda}
\tau^{(\Lambda)}&= \frac{2 v}{3 H^{(\Lambda)}_0  \tanh(v_0)},
\\\label{etaLambda}
a^{(\Lambda)}_0 \eta^{(\Lambda)}&= \frac{2 \cosh(v_0)}{3 H^{(\Lambda)}_0  \left[\sinh(v_0)\right]^{1/3}} \int^{v}_0 \frac{\ud v'}{\left[\sinh(v')\right]^{2/3}},\\
\frac{H^{(\Lambda)}}{H^{(\Lambda)}_0}&= \frac{\tanh(v_0)}{\tanh(v)}.
\end{align}
\end{subequations}
These equations express the standard cosmological behaviour \cite{Hobson} in a form akin to the QB-CDM expansion histories (\ref{exphist}) we previously derived. For $\Lambda$CDM, the time coordinate $v$ runs from the Big Bang $v=0$, to the present day $v=v_0$, and then into the far future $v\to\infty$. As the counterpart to equation (\ref{Omegam}) we can express the matter density parameter as
\begin{align}\label{LambdaOmegam}
\Omega_{\mathrm{m}}^{(\Lambda)}= \left[\cosh(v)\right]^{-2},
\end{align}
which also implies
\begin{align}\label{v0toOmega}
v_0 \equiv \cosh^{-1}\!\left[\left(\Omega_{\mathrm{m}0}^{(\Lambda)}\right)^{-1/2}\right].
\end{align}
The $\Lambda$CDM cosmologies (\ref{LCDM}) are determined by two parameters: $\{H^{(\Lambda)}_0,v_0\}$, or equivalently $\{H^{(\Lambda)}_0,\Omega_{\mathrm{m}0}^{(\Lambda)}\}$. In comparison, the QB-CDM expansion histories (\ref{exphist}) have a single extra parameter: once radiation has been neglected ($\alpha=0$) we are left with $\{H^{\Sh}_0,u_0,\bar{\gamma}\}$.

\subsection{Matching Conditions}\label{matching}
We will explore the full $\{H^{\Sh}_0,u_0,\bar{\gamma}\}$ parameter-space in a future publication, when we test QB-CDM against actual data. Our present aim is more modest: we wish to see how closely QB-CDM can resemble the standard $\Lambda$CDM model of our universe, and hence identify the range of plausible $\bar{\gamma}$. To this end, we shall fix $\{H^{\Sh}_0,u_0\}$ by fiat -- insisting that the QB-CDM universe has the same present-day matter content \begin{subequations}\label{matchcond}
\begin{align}\label{density}
 \rho^{\Sh}_{\mathrm{m}0}=\rho^{(\Lambda)}_{\mathrm{m}0},
\end{align}
and conformal age 
\begin{align}\label{eta0}
a^{\Sh}_0 \eta^{\Sh}_0&=a^{(\Lambda)}_0 \eta^{(\Lambda)}_0,
\end{align} \end{subequations}
as the $\Lambda$CDM universe that best fits the observations from \emph{Planck} \cite{Planck18}: $H_{0}^{(\Lambda)}\cong 67\,\mathrm{km\ s^{-1}Mpc^{-1}}$, $\Omega_{\mathrm{m}0}^{(\Lambda)}\cong 0.31$. Roughly speaking, the first matching condition (\ref{density}) introduces the correct amount of dark matter into QB-CDM, while the second condition (\ref{eta0}) fixes the angular diameter distance of the surface of last scattering. Of course, this \emph{exact} agreement is overly restrictive: in reality, our estimates of $\rho_\mathrm{m0}$ and $a_0\eta_0$ have experimental uncertainty, and are (weakly) model dependent. Nonetheless, it is an interesting exercise to adopt this \emph{common ground} as a simplifying assumption, and then examine how the \emph{other} predictions of QB-CDM differ from $\Lambda$CDM. In this fashion, we will obtain a conservative appraisal of QB-CDM, confident that a better fit can be obtained by relaxing the assumptions above. 

\subsection{Comparison}
Inserting equations (\ref{Omegam}) and (\ref{LambdaOmegam}) into (\ref{density}), and equations (\ref{exphisteta}) and (\ref{etaLambda}) into (\ref{eta0}), we see that the ``matched'' cosmologies obey
\begin{subequations}\label{matcheqs}
\begin{align}\label{u0match}
 \frac{1-u_0}{\sqrt{F^\prime_{\bar{\gamma}}(u_0)}}&=\frac{2}{3 \left[\sinh(v_0)\right]^{1/3}} \int^{v_0}_0 \frac{\ud v}{\left[\sinh(v)\right]^{2/3}},\\\label{H0match}
\frac{H_{0}^{\Sh}}{H_{0}^{(\Lambda)}}&= \frac{- F^{\prime\prime}_{\bar{\gamma}}(u_0)}{\cosh(v_0)\sqrt{F^{\prime}_{\bar{\gamma}}(u_0)}}.
\end{align}
\end{subequations}
Recalling that $v_0$ is set by equation (\ref{v0toOmega}), we can use equations (\ref{u0match}) and (\ref{H0match}) to fix $u_0$ and $H_{0}^{\Sh}$ in turn. The fundamental constant $\bar{\gamma}$ remains as our only free parameter. 

To compare QB-CDM against $\Lambda$CDM, we contrast the expansion rate $H$, and angular diameter distance $D_\mathrm{A}\equiv  a(\eta) \cdot(\eta_0 -\eta)$, as a function of redshift $z\equiv (a_0/a) -1$. Using the expansion histories (\ref{exphist}), (\ref{LCDM}), and the matching equations (\ref{matcheqs}) we obtain
\begin{align}\nonumber
\frac{\delta H}{H}&\equiv \left[\frac{H^\Sh-H^{(\Lambda)}}{H^{(\Lambda)}}\right]_{z^{\Sh}=z^{(\Lambda)}} \\ \label{deltaH}
&=\frac{-\tanh(v_z) \left[F^\prime_{\bar{\gamma}}(u_0)\right]^{3/2}  F^{\prime\prime}_{\bar{\gamma}}(u)  }{\sinh(v_0) \left[F^\prime_{\bar{\gamma}}(u)\right]^{2}} -1,
\end{align}
and 
\begin{align}\nonumber
\frac{\delta D_\mathrm{A}}{D_\mathrm{A}}&\equiv \left[\frac{D^{\Sh}_\mathrm{A}-D^{(\Lambda)}_\mathrm{A}}{D^{(\Lambda)}_\mathrm{A}}\right]_{z^{\Sh}=z^{(\Lambda)}}\\\label{deltaD}
&= \frac{3 \left[\sinh(v_0)\right]^{1/3} (u-u_0)}{2 \sqrt{F^\prime_{\bar{\gamma}}(u_0)} \int^{v_0}_{v_z}\ud v' \left[\sinh(v')\right]^{-2/3}}-1,
\end{align}
where
\begin{align}\label{vz}
v_z\equiv\sinh^{-1}\!\left[\left(\frac{F^\prime_{\bar{\gamma}}(u)}{F^\prime_{\bar{\gamma}}(u_0)}\right)^{3/2} \sinh(v_0)\right],
\end{align}
is the value of $v$ that achieves $z^{(\Lambda)}=z^{\Sh}$. As we move from the Big Bang $u=1$, to the present day $u=u_0$, equations (\ref{deltaH}--\ref{vz}) describe the fractional difference in $H$ and $D_\mathrm{A}$, between QB-CDM and $\Lambda$CDM universes with  the same present-day matter density (\ref{density}) and conformal age (\ref{eta0}), compared at equal redshift.

\begin{figure}
    \centering
\includegraphics[width=0.49\textwidth]{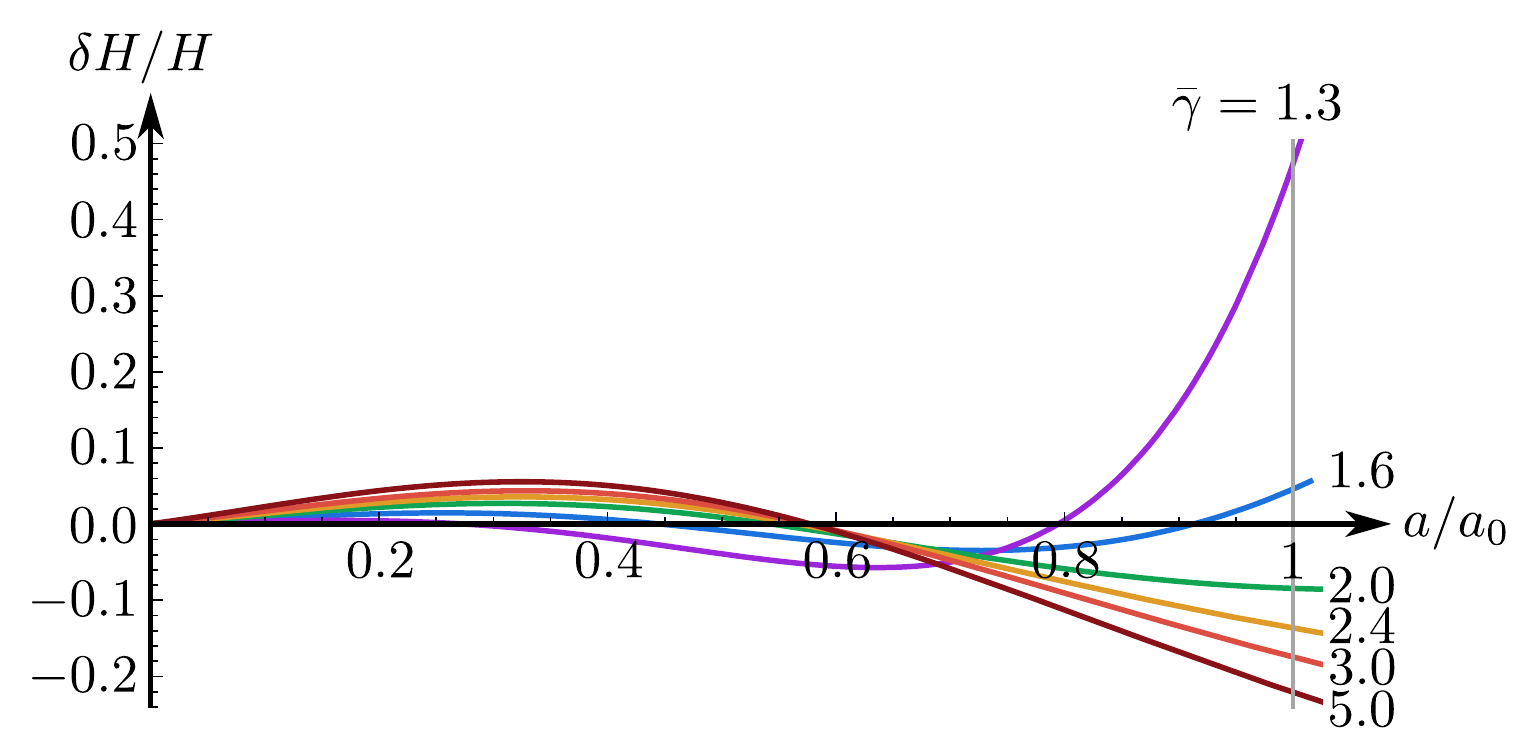}
\includegraphics[width=0.49\textwidth]{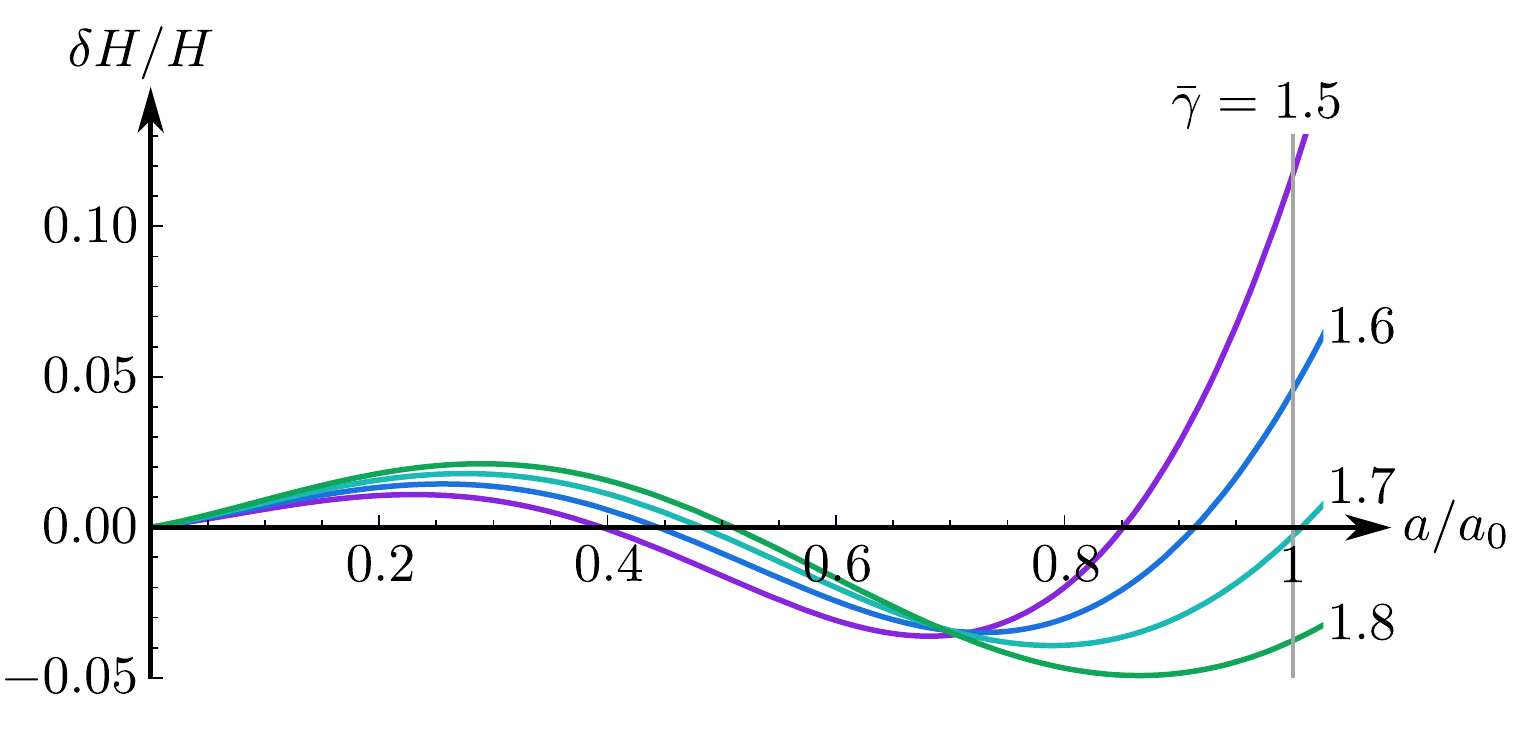}
\includegraphics[width=0.49\textwidth]{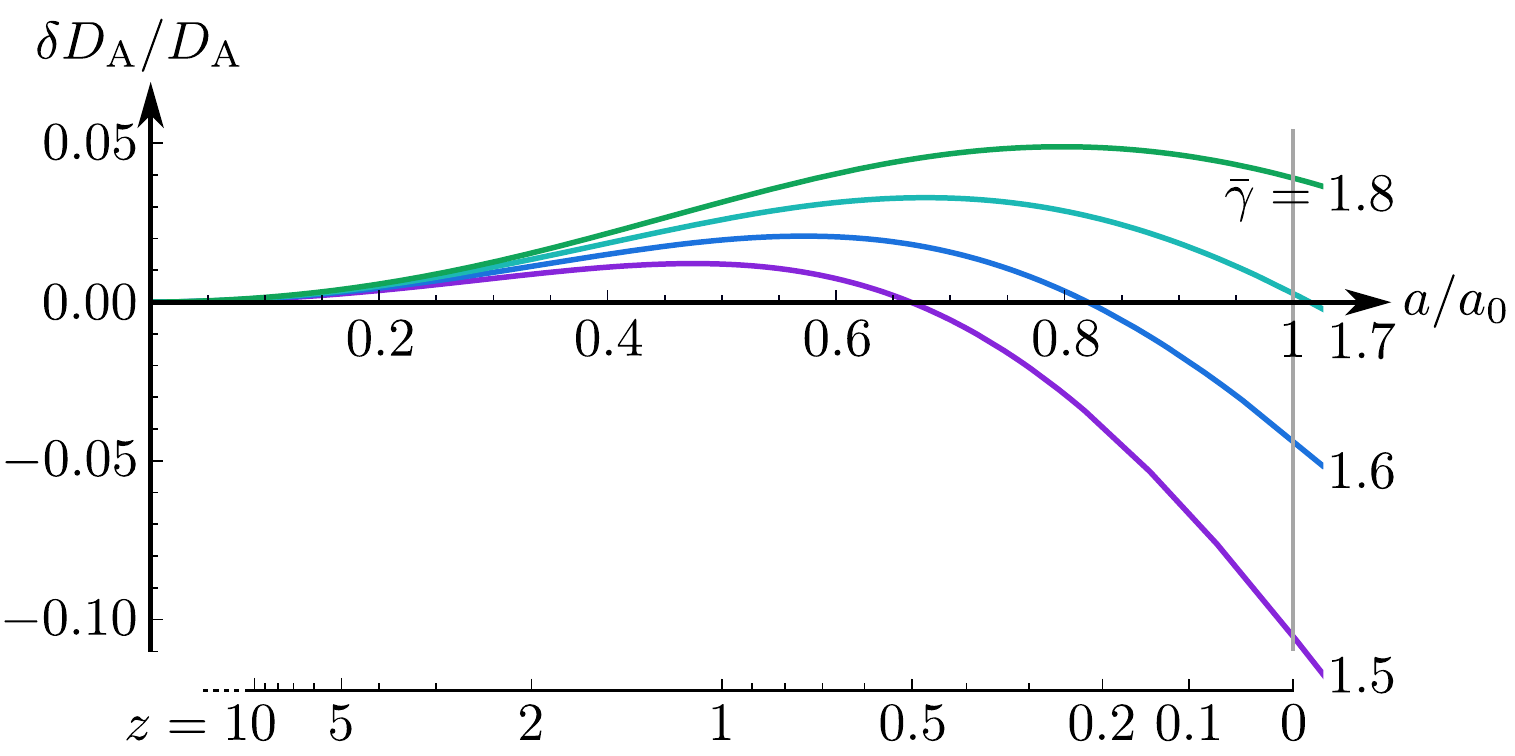}
\includegraphics[width=0.49\textwidth]{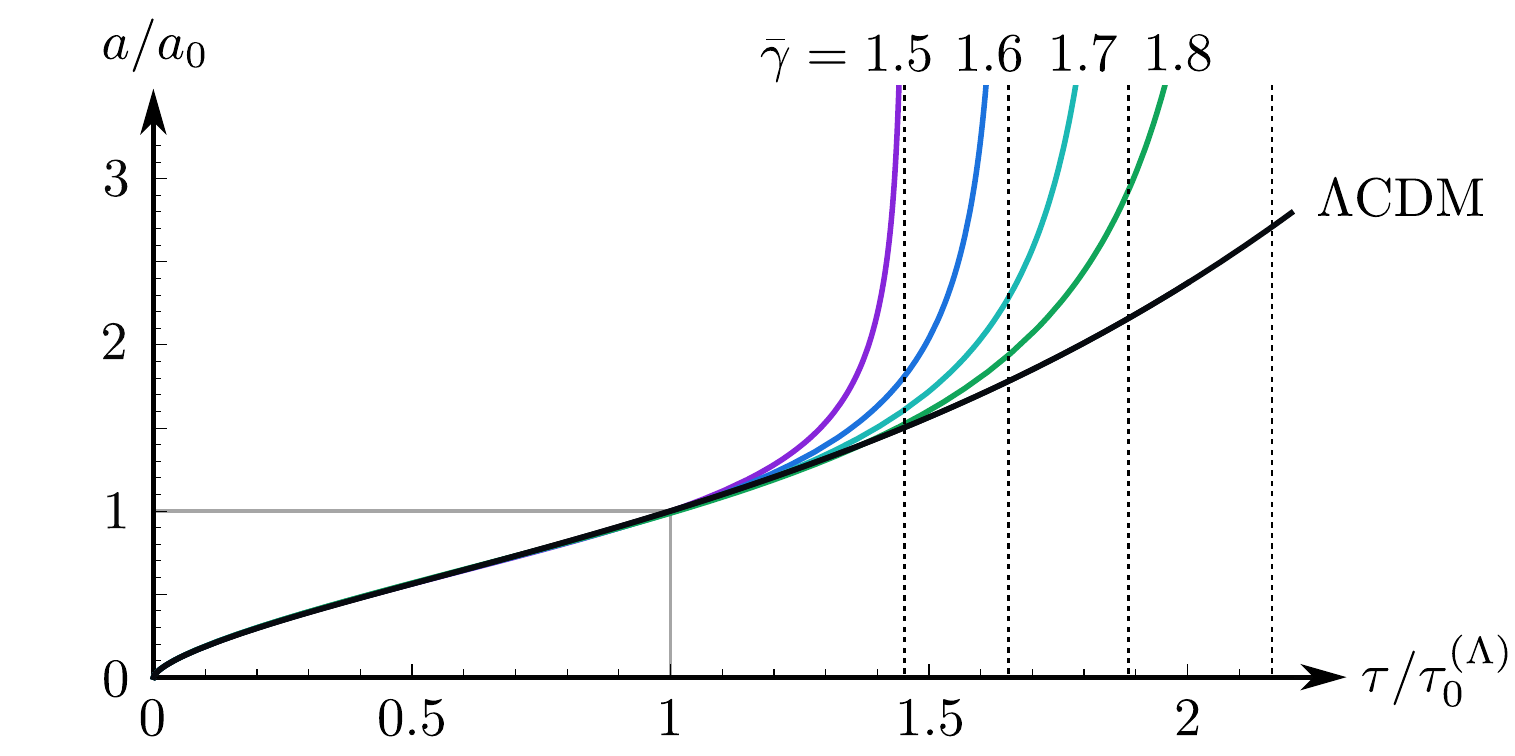}\vspace{-0.41cm}
    \caption{In the plots above, the QB-CDM expansion histories (\ref{exphist}) are compared to $\Lambda$CDM (\ref{LCDM}) with $\Omega^{(\Lambda)}_\mathrm{m0}=0.31$. As explained in section \ref{matching}, the parameters $\{H_0^\Sh,u_0\}$ have been chosen so that the two models are in \emph{exact agreement} over the present-day matter-density (\ref{density}) and conformal age of the universe (\ref{eta0}). The two topmost graphs show the fractional difference in the Hubble expansion rate (\ref{deltaH}) at each redshift: first for the wide range of values $1.3\le\bar{\gamma}\le5$ used in figure \ref{DensityPlot}; then for a small group $\bar{\gamma}\in \{1.5,1.6,1.7,1.8\}$ that agree with $\Lambda$CDM most closely. The third graph depicts the fractional difference in the angular diameter distance (\ref{deltaD})  for the narrow range of $\bar{\gamma}$. Finally, the scale factor is plotted as a function of proper time, for $\bar{\gamma}\in \{1.5,1.6,1.7,1.8\}$ and $\Lambda$CDM. Here, vertical dotted lines indicate the proper times $\bar{\tau}^{\Sh}/\tau_0^{(\Lambda)}\approx\{1.46,1.65,1.89,2.17\}$ at which the respective QB cosmologies undergo a Big Rip.}\label{Plots}
\end{figure}

\subsection{Results}
Using the formulae above, we plot the behaviour of $\delta H/H$, $\delta D_\mathrm{A}/D_\mathrm{A}$, and $a(\tau)$ in figure \ref{Plots}. There are a number of details to notice:
\begin{itemize}
\item There is no $\bar{\gamma}$ for which there is absolute agreement $\delta H(z)=0$ over the entire cosmic history. In general, QB-CDM cannot reproduce $\Lambda$CDM to arbitrary accuracy.\footnote{Sending $\bar{\gamma}\to 1$ ($\Rightarrow \bar{g}\to0$) will remove quantum bias from the semiclassical Friedmann equations (\ref{FEgen}); however, this does not recreate $\Lambda$CDM. There is no cosmological constant in QB-CDM, so this limit corresponds to a classical $\Omega_\mathrm{m}=1$ Einstein-de Sitter universe, which does not accelerate.} The new theory is therefore \emph{falsifiable}.

\item In general, there is close agreement between QB-CDM and $\Lambda$CDM at early times. This occurs for two reasons. Firstly, the primordial equation of state  (\ref{wearly}) ensures that $\rho_\mathrm{eff}$ becomes negligible as $a\to 0$. (For example, $\bar{\gamma}= 1.6$ has $\rho_\mathrm{eff*}/\rho_\mathrm{m*}\approx 8\times 10^{-5}$ at $z_*\approx 1100$.)  Secondly, the matching conditions (\ref{matchcond}) have ``calibrated'' the QB cosmologies such that the limits $\kappa\rho_\mathrm{m 0} = \lim_{a\to0} \{3H^2a^3/a_0^3\}$ and $a_0 \eta_0=\lim_{a\to0}\{D_\mathrm{A}a_0/a\}$ agree \emph{exactly} with $\Lambda$CDM. Consequently, the QB cosmologies considered here will be consistent with observations of the cosmic microwave background (CMB). Indeed, a more realistic treatment would account for the experimental uncertainty in $\rho_0$ and $a_0 \eta_0$: small deviations would be tolerated at $a=0$, allowing closer agreement at late times.

\item At late times, the QB cosmologies diverge from $\Lambda$CDM, and each other. Hence, $\bar{\gamma}$ will be well-constrained by direct measurements of the Hubble constant $H_0$. At present, there is significant tension between the directly measured $H_0=(73.52 \pm 1.62) \mathrm{km}\,{\mathrm{s}}^{-1}{\mathrm{Mpc}}^{-1}$ from standard candles in the local universe \cite{Riess18a,Riess18b}, and $\Lambda$CDM constrained by CMB data: $H^{(\Lambda)}_0=(67.66 \pm 0.42) \mathrm{km}\,{\mathrm{s}}^{-1}{\mathrm{Mpc}}^{-1}$ \cite{Planck18}. As the second plot shows, values near $\bar{\gamma}\approx 1.6$ are able to resolve this tension, generating a deviation $\delta H_0/H_0\approx 5\%$ that would reconcile the present-day expansion rate with observations of the early universe. 

\item At moderate redshift, Baryon Acoustic Oscillations (BAOs) will provide the tightest constraints on QB-CDM. The distances to redshifts near $z=0.5$ have been measured to a precision of roughly $1\%$ \cite{Alam17} and found to be consistent with CMB-constrained $\Lambda$CDM \cite{Planck18}.  Consulting the third plot, we see that QB-CDM with $\bar{\gamma}\approx 1.5$ cannot be distinguished from $\Lambda$CDM by these measurements. Moreover, these values naturally resolve the aforementioned Hubble tension: $\delta H_0/H_0 \approx 10\%$. Once the matching conditions (\ref{matchcond}) are relaxed, the constraint on $\bar{\gamma}$ will loosen -- nonetheless, it appears that current BAO measurements will favour values near $\bar{\gamma}\approx 1.6$, and select QB cosmologies with slightly larger $H_0$ than $\Lambda$CDM.

\item The favoured values $\bar{\gamma}\approx 1.6$ have an effective equation of state (\ref{wdef}) that is phantom $w_\mathrm{eff}<-1$ at late times (\ref{wlate}). We see the consequences (\ref{finaltimes}) of this feature in the fourth plot: the  QB universes end in a Big Rip  at $\bar{\tau}/\tau_0\approx 1.7$. 
\end{itemize}
This brief analysis suggests that current measurements cannot distinguish QB-CDM from $\Lambda$CDM, at least for some values of the parameters $\{H^{\Sh}_0,u_0,\bar{\gamma}\}$. It is therefore unlikely that QB-CDM can be ruled out with present data. In a future paper, I will confront the theory with observational data directly, inferring a posterior distribution for $\{H^{\Sh}_0,u_0,\bar{\gamma}\}$ without using $\Lambda$CDM as a reference model.

\section{Conclusions}
We have motivated and developed a new fundamental theory of cosmic acceleration (Quantum Bias Cosmolgy) that does not require dark energy or modified gravity. Instead, the expansion of the universe is accelerated by a subtle quantum phenomenon \cite{Butcher18a,Butcher18b} that emerges in any system with information capacity $\mathcal{S}$ that depends on a dynamical variable. In general, a quantum correction (\ref{VofS}) induces a \emph{bias} in the behaviour of the system (\ref{meanEOM}) which forces it off its classical trajectory; one accounts for this effect \emph{semiclassically} by including the bias in the action (\ref{semiclasS}). Quantum Bias Cosmology brings this formalism to bear on the universe as a whole, with the cosmological information capacity (\ref{Sholo}) quantified according to the holographic principle (appendix \ref{HoloPrinc}). Once quantum bias (\ref{Vholo}) has been included in the cosmological action (\ref{JGaN}), we arrive at \emph{semiclassical} Friedmann equations (\ref{FEgen}) in which cosmic acceleration (\ref{accn}) arises \emph{automatically}:
\begin{align}\label{accneta}
\frac{1}{a}\frac{\ud^2 a}{\ud \tau^2}=-\frac{\kappa}{6}(\rho +3p)+\frac{2\bar{g}}{a^4}\int^{\eta}_{0} \ud \eta'    \frac{[a(\eta')]^2}{(\bar{\eta}-\eta')^3},
\end{align}
which dependends on the past behaviour of the scale factor. We have solved the semiclassical Friedmann equations for a spatially-flat universe containing matter and radiation (\ref{exphist}). As shown in figure \ref{Plots}, these solutions succeed in reproducing the predictions of $\Lambda$CDM to within the accuracy of current  observations. We conclude that quantum bias provides cosmic acceleration ``for free'', consistent with experiment, as a natural consequence of treating the universe as a holographic quantum system. 

\emph{Free Parameter:} QB-CDM introduces a single unknown dimensionless constant $\bar{\gamma}\equiv\sqrt{4\bar{g}+1}$. For no value of $\bar{\gamma}$ is there an exact match between the predictions of QB-CDM and $\Lambda$CDM, so the new theory is falsifiable. A preliminary analysis (section \ref{Compare}) suggests that CMB+BAO observations favour $\bar{\gamma}\approx 1.6$, generating slightly larger values of $H_0$ than $\Lambda$CDM. (In a subsequent paper, I will determine whether this effect can resolve the well-known tension between local measurements of $H_0$ \cite{Riess18a,Riess18b} and the CMB \cite{Planck18}.) The quantity $\bar{g}=\pi^2 \bar{\mu}^2/\bar{d}$ is set by a numerical filling factor $\bar{\mu}=1/(24 \ln 2 -15) \approx 0.61142$ that accounts for the organisation of holographic information in spacetime (\ref{mubargen}), and a constant $\bar{d}$, defined by equation (\ref{dbardef}), which depends on unknown details of the cosmological configuration space (appendix \ref{BiasDeriv}). In the future, we will investigate whether $\bar{d}$ can be derived from fundamental theory.

\emph{Coincidence:} The favoured values $\bar{\gamma}\approx 1.6$ predict a Big Rip at $\bar{\tau}\approx  1.7 \times \tau_0$. This prediction ameliorates the coincidence problem \cite{Velten14} because there is no longer an infinite future (with $\Omega_\Lambda \cong 1$) where we should expect to find ourselves \cite{Caldwell03,Scherrer05}. Instead, QB-CDM places us at a rather typical point in cosmological history, roughly halfway between the initial singularity $a=0$, and the final singularity $a=\infty$.

\emph{Fine Tuning:} In Quantum Bias Cosmology, the magnitude of cosmic acceleration (\ref{accneta}) is essentially determined by the area of the cosmological event horizon. (This is the reverse of the usual view, wherein $\Lambda$ sets the size of the horizon.) Hence, we can seek to explain the extremely small value $\Lambda_\mathrm{obs}\sim 10^{-122}/\ell_\mathrm{pl}^2$  as the result of some physical process that expands this area at early times.  Inflation is the obvious candidate for such a mechanism, conceivably solving the fine-tuning problem in the same fashion as the flatness problem. I will investigate this possibility in a future publication, when I extend Quantum Bias Cosmology to the very early universe.

\acknowledgments
This research was supported by a research fellowship from the Royal Commission for the Exhibition for 1851, and by the Institute for Astronomy at the University of Edinburgh. The author also wishes to thank John Peacock, Lucas Lombriser, Alex Hall, Yan-Chuan Cai, and Joe Zuntz for helpful discussions.

\appendix

\section{\\DISCARDED DEGREES OF FREEDOM}\label{BiasDeriv}
Here we summarise the derivation of the quantum bias formula
\begin{align}\label{VofSApp}
\!\!\Delta V_\mathrm{eff}=\frac{\hbar^2}{8m}\!\left[\!\left(\!1- 4\xi \frac{d+1}{d}\!\right)\!\!\left(\partial_x \mathcal{S}\right)^2\!+ 2(1-4\xi)\partial_x^2 \mathcal{S}\right]\!\!,\!\!
\end{align}
and briefly discuss how this result might be generalised.

In the first paper of this series \cite{Butcher18a}, equation (\ref{VofSApp}) is derived by modelling the full configuration space of the classical system (\ref{reducedAction}) as a warped manifold:
\begin{align}\label{DoFmetric}
 \ud s^2 &= \ud x^2 + e^{2\mathcal{S}(x)/d} g_{ij}(\varphi)\ud \varphi^i \ud \varphi^j,
\end{align}
so that the discarded variables $\varphi\in \mathcal{M}_\varphi$ cover a closed $d$-dimensional submanifold of physical volume $\text{vol}[\mathcal{M}_\varphi]\propto \exp[\mathcal{S}(x)]$. Once the system is quantised (and UV regularised) the discarded Hilbert subspace $\mathcal{H}_\varphi$ then has $\dim[\mathcal{H}_\varphi]\propto \text{vol}[\mathcal{M}_\varphi]\propto \exp[\mathcal{S}(x)]$ as required. (The constants of proportionality, and the UV regulator, drop out of the final result). The quantised system is evolved according to a covariant Schr\"odinger equation over the curved configuration space (\ref{DoFmetric}); this equation is unique up to a curvature-coupling term with constant coefficient $\xi\in \mathbb{R}$, the only significant quantisation ambiguity. Once $\mathcal{H}_\varphi$ is discarded, one arrives at a Schr\"odinger equation for the $x$ observable alone; therein, one finds the potential to be $V_\mathrm{cl} + \Delta V_\mathrm{eff}$, differening from the classical system (\ref{reducedAction}) by the above quantum correction (\ref{VofSApp}). Besides the constants $\xi$ and $d$, this result is completely independent of the internal geometry  of the discarded configuration space $g_{ij}(\varphi)$. In this sense, equation (\ref{VofSApp}) generically captures the effect of a dynamic information capacity $\mathcal{S}(x)$. 

The path integral approach \cite{Butcher18b} allows us extend this reasoning to discarded degrees of freedom with a \emph{history-dependent} information capacity
\begin{align}\label{Sint}
\mathcal{S}=\mathcal{S}\!\left(x,\int^t \ud t' f(x(t'))\right),
\end{align}
which includes $\mathcal{S}=\mathcal{S}(x,t)$ as the special case  $f=1$. The formula (\ref{VofSApp}) is unchanged by this generalisation, with the $\partial_x$ derivatives acting only on the first argument of $\mathcal{S}$. (In particular, unitary evolution ensures that $\partial_t \mathcal{S}$ terms do not appear.) The formula (\ref{VofSApp}) is therefore sufficiently powerful to capture the most general form of cosmological information capacity $\mathcal{S}=\mathcal{S}(a,\int^\eta \ud \eta' f(a(\eta')))$ considered in this paper.

Beyond the history-dependent extension (\ref{Sint}) of the warped configuration space (\ref{DoFmetric}) there does not appear much to be gained. The warped metric can obviously be generalised; however, these nonminimal models typically introduce new functions $\lambda(x)$ that have no relation to the discarded information capacity $\mathcal{S}(x)$. Without a fundamental motivation for these new functions, and some physical principles to constrain them, there is little reason to explore such models in detail. 

As an alternative approach, we can  ignore the structure of configuration space entirely, and simply write down the most general $\Delta V_\mathrm{eff}$ that can be formed from $\{\hbar, m, \mathcal{S}\}$ and $\partial_x$ derivatives. With this method, dimensional considerations restrict us to
\begin{align}\label{VGen}
 \Delta V_\mathrm{eff}=\frac{\hbar^2}{m}\sum_k\left[A_k \mathcal{S}^k (\partial_x\mathcal{S})^2 +B_k \mathcal{S}^k \partial^2_x\mathcal{S}\right],
\end{align}
where $\{A_k,B_k: k \in \mathbb{Q}\}$ are a set of unknown dimensionless constants. But notice: we can always redefine our system (\ref{reducedAction}) by including \emph{irrelevant} degrees of freedom, i.e.\ discarded variables $\varphi'$ that are completely independent of $x$ and $\varphi$. These redefinitions send $\mathcal{S}\to \mathcal{S}+\text{const}$, but cannot affect the behaviour of $x$; hence, they cannot cause more than a shift $\Delta V_\mathrm{eff}\to \Delta V_\mathrm{eff}+\text{const}$. This argument forces us to set $A_k=B_k=0$ for all $k\ne 0$, reducing our general construction (\ref{VGen}) to the standard form (\ref{VofSApp}). The net effect of this abstraction is to replace $(\xi,d)\in \mathbb{R}\times \mathbb{N}$ with a slightly larger parameter space $(A_0,B_0)\in \mathbb{R}^2$ that has no obvious  physical interpretation. As far as the conclusions of this paper are concerned, this generality is equivalent to allowing $d$ to take noninteger values.

To see how $d\not\in \mathbb{N}$ might arise concretely, consider a separable discarded configuration space $\mathcal{M}_\varphi=\mathcal{M}_{\varphi(1)}\times\ldots\times\mathcal{M}_{\varphi(N)}$, where each ($d_n$-dimensional) submanifold $\mathcal{M}_{\varphi(n)}$ scales at a different rate:
\begin{align}\label{DoFmetricShape}
 \ud s^2 &= \ud x^2 + \sum_{n=1}^N e^{2\alpha_n\mathcal{S}(x)/d_n} g^{(n)}_{ij}(\varphi_{(n)})\ud \varphi_{(n)}^i \ud \varphi^j_{(n)}.
\end{align}
Here, we have introduced $N$ free parameters $\alpha_n\in \mathbb{R}$, but no free functions. (In fact, there are only $N-1$ free parameters: we need $\sum_{n} \alpha_n=1$ to ensure $\text{vol}[\mathcal{M}_\varphi]\propto \exp[\mathcal{S}(x)]$.) In this model, the discarded space not only changes size as a function of $x$, it also changes \emph{shape}. Rerunning the derivation \cite{Butcher18a}, one finds that the only modification to equation (\ref{VofSApp}) is the replacement
\begin{align}\label{drep}
\left(1-4\xi \frac{d+1}{d}\right) \to  \sum_{n=1}^N \alpha_n^2\left(1-4\xi \frac{d_n+1}{d_n}\right),
\end{align}
in the first term. For the cosmologically preferred value $\xi=1/4$ (see appendix \ref{Gauge}) the replacement (\ref{drep}) becomes
\begin{align}\label{dnonint}
\frac{1}{d}\to \sum_{n=1}^N \frac{ \alpha_n^2}{d_n} \in \mathbb{R}^+,
\end{align}
which can be realised in equation (\ref{VofSApp}) by allowing $d$ to take positive noninteger values.

\section{\\NEW VARIABLES AND GAUGE INVARIANCE}\label{Gauge}
In this appendix, we examine the extent to which cosmological quantum bias (\ref{VofSa}) is consistent with two key symmetries of the classical theory: (i) the gauge freedom of the time coordinate, and (ii) our ability to redefine the dynamical variable $a=f(\widetilde{a})$. To keep this discussion self-contained, let us briefly summarise the process by which  the semiclassical action (\ref{JGaN}) is derived. 

Starting with the metric
\begin{align}\label{A.FRWmetric}
\ud s^2 = [a(t)]^2\left(-[N(t)]^2\ud t^2 + \ud \chi^2 + [r_k(\chi)]^2\ud \Omega^2\right),
\end{align}
we first obtain the classical gravitational action (\ref{IGaN}):
\begin{align}\label{A.IGaN}
\mathcal{I}_\mathrm{G}[a(t),N(t)]= \frac{3\mathscr{V}_*}{\kappa}\int^{t_+}_{t_-} \ud t \left[ -\frac{\dot{a}^2}{N} + k N a^2  \right].
\end{align}
The conformal time coordinate $\eta=\eta(t)$, defined by
\begin{align}\label{A.etadef}
\ud \eta&=  N \ud t, & \eta_\pm&\equiv \eta(t_\pm),
\end{align}
then allows us to write the action (\ref{A.IGaN}) in canonical form
\begin{align}\label{A.IGa}
\mathcal{I}_\mathrm{G}[a(\eta)]= \frac{3\mathscr{V}_*}{\kappa}\int^{\eta_+}_{\eta_-} \ud \eta \left[ -\left(\frac{\ud{a}}{\ud\eta}\right)^2 + k a^2 \right].
\end{align}
Comparing this action with the reference (\ref{reducedAction}), we formally identified $x\to a$, $t\to \eta$, $m\to -6\mathscr{V}_*/\kappa$; hence, the quantum bias (\ref{VofS}) becomes (\ref{VofSa}), and the semiclassical action (\ref{semiclasS}) is
\begin{align}\nonumber
\mathcal{J}_\mathrm{G}[a(\eta)]&= \frac{3\mathscr{V}_*}{\kappa}\int^{\eta_+}_{\eta_-}\!\! \ud \eta \left[ -\left(\frac{\ud{a}}{\ud\eta}\right)^2 + k a^2  
\right.\\\label{A.JGa}
&\quad {}+Q_1 \left(\partial_a \mathcal{S}\right)^2 + Q_2 \partial_a^2\mathcal{S}\Bigg],
\end{align}
where $\mathcal{S}=\mathcal{S}(a,\eta)$ is the information capacity of the discarded degrees of freedom, and
\begin{align}\label{Cdef}\bs
Q_1&\equiv \frac{4 \pi^2 \ell_\mathrm{pl}^4}{9\mathscr{V}^2_*}\left(1- 4\xi \frac{d+1}{d}\right),\\
Q_2&\equiv \frac{8 \pi^2 \ell_\mathrm{pl}^4}{9\mathscr{V}^2_*}\left(1- 4\xi\right)\es
\end{align}
depend on the unknown constants $\xi$ and $d$. Finally, we re-express the semiclassical action (\ref{A.JGa}) in terms of the generic time coordinate $t$,
\begin{align}\nonumber
\mathcal{J}_\mathrm{G}[a(t),N(t)]&= \frac{3\mathscr{V}_*}{\kappa}\int^{t_+}_{t_-}\ud t \left[ -\frac{\dot{a}^2}{N} + k N a^2  \right.
\\\label{A.JGaN}
&\quad{}+N \left(Q_1 \left(\partial_a \mathcal{S}\right)^2 + Q_2 \partial_a^2\mathcal{S}\right)\bigg],
\end{align}
so that the semiclassical Friedmann equations (\ref{FEgen}) can be obtained by variations $\delta a(t)$, $\delta N(t)$.

For the present discussion, the critical step above is the selection of $\eta$ as the time coordinate that renders $\mathcal{I}_G$ in the canonical form (\ref{A.IGa}). At first glance, it appears that $\eta$ is the only such coordinate that can achieve this goal, allowing us to make contact with the quantum theory of section \ref{DoFs}. However, suppose we define the scale factor using an invertible differentiable function $f$,
\begin{align}\label{a=f}
a= f\!\left(\widetilde{a}(t)\right),
\end{align}
and consider $\widetilde{a}(t)$ and $N(t)$ as our  new dynamical variables. Then the classical action (\ref{A.IGaN}) becomes
\begin{align}
\widetilde{\mathcal{I}}_\mathrm{G}[\widetilde{a}(t),N(t)]&\equiv \mathcal{I}_\mathrm{G}[f(\widetilde{a}(t)),N(t)]\\ \nonumber
&= \frac{3\mathscr{V}_*}{\kappa}\!\int^{t_+}_{t_-}\! \ud t\! \left[- \frac{\dot{\widetilde{a}}{}^2}{N}\left[f'\!\left(\widetilde{a}\right)\right]^2 +k  N  \left[f\!\left(\widetilde{a}\right)\right]^2 \right]\!,
\end{align}
which takes on canonical form
\begin{align}\nonumber
&\widetilde{\mathcal{I}}_\mathrm{G}[\widetilde{a}(\widetilde{\eta})]=
\\\label{IGat}
&\qquad\frac{3\mathscr{V}_*}{\kappa}\int^{\widetilde{\eta}_+}_{\widetilde{\eta}_-} \ud \widetilde{\eta} \left[ -\left( \frac{\ud \widetilde{a}}{\ud \widetilde{\eta}}\right)^2 + k  \left[{f'\!\left(\widetilde{a}\right)\! f\!\left(\widetilde{a}\right)}\right]^2 \right],
\end{align}
when we use a new time coordinate $\widetilde{\eta}=\widetilde{\eta}(t)$, with
\begin{align}\label{tetadef}
\ud \widetilde{\eta}&=  \left[f'\!\left(\widetilde{a}\right)\right]^{-2}N \ud t, & \widetilde{\eta}_\pm&\equiv \widetilde{\eta}(t_\pm),
\end{align}
as its defining equations.

As far as the classical theory is concerned, the pair $(\widetilde{a},\widetilde{\eta})$ stand on the same footing as $(a,\eta)$. General covariance regards $\eta$ and $\widetilde{\eta}$ as equally valid coordinates, and  there is no reason \emph{a priori} that the spacetime (\ref{A.FRWmetric}) should be parametrised by $a$, rather than $\widetilde{a}=1/a$ or $\widetilde{a}=a^2$, say. Furthermore, since $\widetilde{\mathcal{I}}_\mathrm{G}[\widetilde{a}(\widetilde{\eta})]$ has the canonical form (\ref{reducedAction}) we are free to apply the quantum theory asserted in section \ref{DoFs}, and hence derive a new semiclassical action $\widetilde{\mathcal{J}}_\mathrm{G}[\widetilde{a}(\widetilde{\eta})]$. The question is -- will this $\widetilde{\mathcal{J}}_\mathrm{G}$ agree with the semiclassical action (\ref{A.JGaN}) derived with our original variables? In other words: does the $(\widetilde{a},\widetilde{\eta})\leftrightarrow(a,\eta)$ equivalence survive the quantum correction?

To answer this question, we shall calculate $\widetilde{\mathcal{J}}_\mathrm{G}$ explicitly, and see how it differs from $\mathcal{J}_\mathrm{G}$. Exactly as before, we compare the classical action (\ref{IGat}) to the standard (\ref{reducedAction}) and see that we must now identify $x\to \widetilde{a}$, $t\to \widetilde{\eta}$, and $m\to -6\mathscr{V}_*/\kappa$. Quantum bias (\ref{VofS}) therefore transforms the classical action (\ref{IGat}) into the following semiclassical action: 
\begin{widetext}
\begin{align}\label{JGat}
\widetilde{\mathcal{J}}_\mathrm{G}[\widetilde{a}(\widetilde{\eta})]&=\frac{3\mathscr{V}_*}{\kappa}\int^{\widetilde{\eta}_+}_{\widetilde{\eta}_-} \ud \widetilde{\eta} \left[ -\left( \frac{\ud \widetilde{a}}{\ud \widetilde{\eta}}\right)^2 + k  \left[{f'\!\left(\widetilde{a}\right)\! f\!\left(\widetilde{a}\right)}\right]^2  +\widetilde{Q}_1 \big(\partial_{\tilde{a}} \mathcal{S}\big)^2 + \widetilde{Q}_2 \partial^2_{\tilde{a}}\mathcal{S} \right],
\end{align}
with $\widetilde{Q}_1$ and $\widetilde{Q}_1$ defined by (\ref{Cdef}) but allowing the unknowns to take new values $(\widetilde{\xi},\widetilde{d})$ for the sake of generality. To evaluate the last two terms in (\ref{JGat}) we will need to write the discarded information capacity $\mathcal{S}(a,\eta)$ as a function of our new variables $(\widetilde{a},\widetilde{\eta})$. This is achieved by noting that (\ref{A.etadef}) and (\ref{tetadef}) imply
\begin{align}\label{et=}
\eta(\widetilde{\eta}) =\eta_-+\int^{\widetilde{\eta}}_{\widetilde{\eta}_-} \ud \widetilde{\eta}' \left[f'\!\left(\widetilde{a}(\widetilde{\eta}')\right)\right]^{2}, \qquad \Rightarrow\qquad
\mathcal{S}(a,\eta)= \mathcal{S}\left(f(\widetilde{a}),\eta_-+\int^{\widetilde{\eta}}_{\widetilde{\eta}_-} \ud \widetilde{\eta}' \left[f'\!\left(\widetilde{a}(\widetilde{\eta}')\right)\right]^{2}\right).
\end{align}
In terms of $(\widetilde{a},\widetilde{\eta})$, the information capacity $\mathcal{S}$ is history dependent (\ref{Sint}) so the path integral construction \cite{Butcher18b} ensures the validity of (\ref{JGat}) with the $\partial_{\tilde{a}}$ derivatives acting on the first argument of $\mathcal{S}$ only. Thus, for the purposes of calculating (\ref{JGat}) we have
\begin{align}\label{dStilde}
\partial_{\tilde{a}} \mathcal{S}&= f'\!\left(\widetilde{a}\right)\partial_a \mathcal{S},&
\partial_{\tilde{a}}^2 \mathcal{S}&= \left[f'\!\left(\widetilde{a}\right)\right]^{2}\partial_a^2 \mathcal{S} + f''\!\left(\widetilde{a}\right)\partial_a \mathcal{S}.
\end{align}
Inserting these formulae into equation (\ref{JGat}) we obtain 
\begin{align}\label{JGanew}
\widetilde{\mathcal{J}}_\mathrm{G}[\widetilde{a}(\widetilde{\eta})]&=\frac{3\mathscr{V}_*}{\kappa}\int^{\widetilde{\eta}_+}_{\widetilde{\eta}_-} \ud \widetilde{\eta} \left[ -\left( \frac{\ud \widetilde{a}}{\ud \widetilde{\eta}}\right)^2 + k  \left[{f'\!\left(\widetilde{a}\right)\! f\!\left(\widetilde{a}\right)}\right]^2 +\widetilde{Q}_1 \left[f'\!\left(\widetilde{a}\right)\right]^{2}\left(\partial_a \mathcal{S}\right)^2 + \widetilde{Q}_2 \left(\left[f'\!\left(\widetilde{a}\right)\right]^{2}\partial_a^2 \mathcal{S} + f''\!\left(\widetilde{a}\right)\partial_a \mathcal{S}\right) \right],
\end{align}
as our new semiclassical action.

We are now in a position to ``close the loop'' of this calculation, and return to our original dynamical variables $a(t)$ and $N(t)$. We first use (\ref{tetadef}) to write (\ref{JGanew}) as an integral over $t$,
\begin{align} \label{Jnew}
\widetilde{\mathcal{J}}_\mathrm{G}[\widetilde{a}(t),N(t)]&= \frac{3\mathscr{V}_*}{\kappa}\int^{t_+}_{t_-} \ud t \left[ -\frac{\dot{\widetilde{a}}{}^2}{N}\left[f'\!\left(\widetilde{a}\right)\right]^2 +k  N  \left[f\!\left(\widetilde{a}\right)\right]^2  +\widetilde{Q}_1 N \left(\partial_a \mathcal{S}\right)^2 + \widetilde{Q}_2 N \left(\partial_a^2 \mathcal{S} + \frac{f''\!\left(\widetilde{a}\right)}{\left[f'\!\left(\widetilde{a}\right)\right]^{2}}\partial_a \mathcal{S}\right) \right],\
\end{align}
and then invert (\ref{a=f}) to express everything as a function of $a(t)$:
\begin{align}
\widetilde{\mathcal{J}}_\mathrm{G}[f^{-1}(a(t)),N(t)]&= \frac{3\mathscr{V}_*}{\kappa}\int^{t_+}_{t_-}\ud t \Bigg[- \frac{\dot{a}^2}{N} + k N a^2 +\widetilde{Q}_1 N \left(\partial_a \mathcal{S}\right)^2 + \widetilde{Q}_2 N \left(\partial_a^2 \mathcal{S} + \frac{f''\!\left(f^{-1}(a)\right)}{\left[f'\!\left(f^{-1}(a)\right)\right]^{2}}\partial_a \mathcal{S}\right) \Bigg].
\end{align}
Comparing this with our original semiclassical action (\ref{A.JGaN}) we see that the $(\widetilde{a},\widetilde{\eta})$ approach has altered our result by
\begin{align}\label{DeltaJG}
\Delta \mathcal{J}_\mathrm{G}\equiv\widetilde{\mathcal{J}}_\mathrm{G}-\mathcal{J}_\mathrm{G}= \frac{3\mathscr{V}_*}{\kappa}\int^{t_+}_{t_-} \ud t\, N\Bigg[\left(\widetilde{Q}_1 -Q_1 \right) \left(\partial_a \mathcal{S}\right)^2 + \left(\widetilde{Q}_2 -Q_2 \right) \partial_a^2 \mathcal{S}+  \widetilde{Q}_2\frac{f''\!\left(f^{-1}(a)\right)}{\left[f'\!\left(f^{-1}(a)\right)\right]^{2}}\partial_a \mathcal{S}\Bigg].
\end{align}
\end{widetext}
Notice that there are no $t$--derivatives in the integrand, so $\Delta \mathcal{J}_\mathrm{G}$ contains no surface terms. Hence, $\widetilde{\mathcal{J}}_\mathrm{G}$ and $\mathcal{J}_\mathrm{G}$ will generate identical semiclassical behaviour if and only if $\Delta \mathcal{J}_\mathrm{G}=0$. Assuming that $\partial_a \mathcal{S}$ and $\partial_a^2 \mathcal{S}$ are not identically zero, then the only way to achieve $\Delta \mathcal{J}_\mathrm{G}=0$ for all $f$ is to set $\widetilde{Q}_1=Q_1$ and $\widetilde{Q}_2=Q_2=0$.\footnote{\emph{Proof:} Given that $\partial_a\mathcal{S}\ne 0$, each choice of $f$ will alter the way the last term of (\ref{DeltaJG}) depends on $a$; in contrast, the other terms can only depend on $f$ through the constants $\widetilde{Q}_1$ and $\widetilde{Q}_2$, and this does not change their $a$-dependence. Hence, $\Delta \mathcal{J}_\mathrm{G}$ can only vanish for all $f$ if this last term vanishes, meaning $\widetilde{Q}_2=0$ is required. But then $\Delta \mathcal{J}_\mathrm{G}$ can only depend on $f$ through the first term $\widetilde{Q}_1 N (\partial_a \mathcal{S})^2$, and as we need $\Delta \mathcal{J}_\mathrm{G}=0$ independent of $f$, we must have $\widetilde{Q}_1$ independent of $f$ also. But then consistency with the trivial case $f(\widetilde{a})=\widetilde{a}$ reveals that $\widetilde{Q}_1=Q_1$. This leaves $-Q_2 N \partial_a^2 \mathcal{S}$ as the only term in the integrand of (\ref{DeltaJG}), so $Q_2=0$ is required also.} %
Consulting (\ref{Cdef}) we see that this is equivalent to
\begin{align}
\xi &=\widetilde{\xi}=1/4,& d&=\widetilde{d}.
\end{align}
We conclude that quantum bias (\ref{VofSa}) is consistent with (i) the gauge invariance of $t$, and (ii) arbitrary redefinitions of the dynamical variable $a=f(\widetilde{a})$, if and only if $d$ is independent of $f$, and $\xi=1/4$.

\section{\\THE HOLOGRAPHIC UNIVERSE}\label{HoloPrinc}
Here we derive the holographic formula (\ref{Sholo}) that quantifies the information capacity of a comoving volume (\ref{Mdef}) of the FRW universe (\ref{FRWmetric}). We begin with a brief review of the holographic principle.

\subsection{The Holographic Principle}
As Bekenstein first realised \cite{Beken81}, the maximum entropy (or information) of a system is not set by its \emph{volume}, but by the \emph{area} of an enclosing surface. This understanding arose from the study of black hole thermodynamics \cite{Beken72,Beken73,Beken74, Bardeen73,Bardeen73,Hawking74,Hawking75}, culminating in the Bekenstein-Hawking formula
\begin{align}\label{SBH}
S_\mathrm{BH}= \frac{A}{4\ell_\mathrm{pl}^2},
\end{align}
for the entropy of a black hole, $A$ being the area of its event horizon. Roughly speaking, $S_\mathrm{BH}$ is the maximum entropy that can ever be stored within a region enclosed by a surface of area $A$. (If this upper bound were ever violated $S>S_\mathrm{BH}$, we could always send energy in through the surface until the region became a black hole. This process would lower the entropy $S\to S_\mathrm{BH}$, and hence violate the second law of thermodynamics.) This idea was given a precise and general formulation by Bousso \cite{Bousso99a} as the covariant entropy bound:
\begin{align}\label{Bbound}
S[\mathcal{L}]\le \frac{A[\mathcal{B}]}{4\ell_\mathrm{pl}^2}.
\end{align}
Here, $A[\mathcal{B}]$ is the area of an arbitrary two-dimensional spacelike surface $\mathcal{B}$, and $S[\mathcal{L}]$ is the entropy on a \emph{lightsheet} $\mathcal{L}$ (a hypersurface of null geodesics with nonpositive expansion) that originates orthogonal to $\mathcal{B}$. Because $\mathcal{L}$ can be past-directed or future-directed, Bousso's bound (\ref{Bbound}) is symmetric under time-reversal, and cannot be understood as a purely thermodynamical statement \cite{Bousso99b}. We are therefore compelled to interpret (\ref{Bbound}) as arising from the number of independent microscopic degrees of freedom present in nature. 

The holographic principle \cite{Hooft93, Susskind95, Bousso99b, Bousso02} elevates these insights to a guiding rule for quantum gravity. At the most basic level, it asserts that the entire (quantum-gravity) state on $\mathcal{L}$ can always be encoded on $\mathcal{B}$, using qubits that occupy an area no less than $\delta A =4(\ln 2)  \ell_\mathrm{pl}^2$. In other words, the states of $\mathcal{L}$ live in a Hilbert space $\mathcal{H}_\mathcal{L}$ of dimension $\dim[\mathcal{H}_\mathcal{L}]\le 2^{A [\mathcal{B}]/\delta A}$, meaning that $\mathcal{L}$ has information capacity
\begin{align}\label{InfoBound}
\mathcal{S}[\mathcal{L}]\equiv \ln\left(\dim[\mathcal{H}_\mathcal{L}]\right)\le \frac{A [\mathcal{B}]}{4\ell_\mathrm{pl}^2}.
\end{align}
Under this premise, the entropy bound (\ref{Bbound}) becomes trivial, because the entropy of a system can never exceed its information capacity: $S\le \mathcal{S}$. 

For this article, we will not need to know \emph{how} the states of $\mathcal{L}$ are encoded on $\mathcal{B}$, nor the process by which three-dimensional physics is expected to emerge from a two-dimensional theory \cite{Maldacena97}. Nonetheless, it is sometimes useful to fix the geometry of $\mathcal{B}$, and explore the range of $\mathcal{L}$-states that can be encoded. For instance, let us consider the case where $\mathcal{B}$ has the geometry of a sphere. Within a semiclassical approximation, each state encoded on $\mathcal{B}$ should determine the geometry and matter content of a lightsheet $\mathcal{L}$ that extends into the interior of $\mathcal{B}$. Now, some of these states will correspond to the interior of a Schwarzschild black hole with event horizon at $\mathcal{B}$; indeed, the Bekenstein-Hawking entropy (\ref{SBH}) must count all such states. Comparing this entropy to (\ref{InfoBound}), and recalling that $S\le \mathcal{S}$, we conclude that the information capacity bound is \emph{saturated},
\begin{align}\label{InfoSat}
\mathcal{S}[\mathcal{L}]= \frac{A [\mathcal{B}]}{4\ell_\mathrm{pl}^2},
\end{align} 
whenever $\mathcal{B}$ is spherical.\footnote{Strictly speaking, $\mathcal{S}[\mathcal{L}]$ must be slightly larger than $S_\mathrm{BH}$, because $S_\mathrm{BH}$ only measures the subspace of $\mathcal{H}_\mathcal{L}$ spanned by states that correspond to the interior of a Schwarzschild black hole with event horizon at $\mathcal{B}$. Indeed, we should have $\mathcal{S}[\mathcal{L}]= S_\mathrm{BH} + I_\mathrm{BH}$, where $I_\mathrm{BH}>0$ is the amount of information conveyed by the statement ``$\mathcal{B}$ is the event horizon of a Schwarzschild black hole''. This information is simply the \emph{macrostate} of $\mathcal{L}$, including its total mass $M=2\sqrt{\pi A[\mathcal{B}]}/\kappa$ and angular momentum $J=0$. However, (\ref{SBH}) and (\ref{InfoSat}) suggest that $\mathcal{S}[\mathcal{L}]=S_\mathrm{BH}$, i.e.\ that $I_\mathrm{BH}$ is negligible within the semiclassical approximation, $A[\mathcal{B}]\gg\ell_\mathrm{pl}^2$.  This comes about because the smallest quantum of energy that can be confined to $\mathcal{B}$ is a massless particle of wavelength $\lambda \sim O(\sqrt{A[\mathcal{B}]})$. Hence $\mathcal{H}_\mathcal{L}$ must have a discrete energy spectrum with minimum spacing $\delta M\sim O(\hbar /\sqrt{A[\mathcal{B}]}\,)$. The macrostate information will then be $I_\mathrm{BH}\sim O(\ln(M/\delta M)) \sim  O(\ln(A[\mathcal{B}]/\ell_\mathrm{pl}^2))\ll \mathcal{S}[\mathcal{L}]$, as claimed.} This is the key holographic result that will allow us to quantify the information capacity of a homogenous, isotropic, expanding universe.

\subsection{Holograms for Cosmology}
To apply equation (\ref{InfoSat}) to cosmology, we require a family of (spherical) surfaces $\mathcal{B}$, whose lightsheets $\mathcal{L}$ cover the entire FRW spacetime (\ref{FRWmetric}). It is natural to insist that the ``holograms'' $(\mathcal{B},\mathcal{L})$ respect the symmetries of the metric; hence, each surface $\mathcal{B}$ should indeed be spherical, and must lie on some hypersurface of simultaneity $t=\text{const}$. To complete our universal covering, we need to specify (i) the size of each $\mathcal{B}$, (ii) whether the $\mathcal{L}$ are directed into the past or future, and (iii) how the holograms $(\mathcal{B},\mathcal{L})$ are arranged in spacetime.

Let us start by imagining we have selected a hologram $(\mathcal{B},\mathcal{L})$ as a candidate for our universal covering. Now suppose we can construct a larger hologram $(\mathcal{B}',\mathcal{L}')$ that completely engulfs our candidate: $\mathcal{L}' \supset  \mathcal{L}$. In principle, equation (\ref{InfoSat}) should apply to both holograms. However, $(\mathcal{B}',\mathcal{L}')$ is clearly a more fundamental description, as it contains $(\mathcal{B},\mathcal{L})$ as a subsystem. We should therefore discard the candidate $(\mathcal{B},\mathcal{L})$ and use the larger hologram $(\mathcal{B}',\mathcal{L}')$ instead. By this logic, our universal covering must be composed of holograms that are \emph{maximal}, i.e.\ those for which no such superset holograms exist.

\begin{figure}\centering
\includegraphics[scale=.9495]{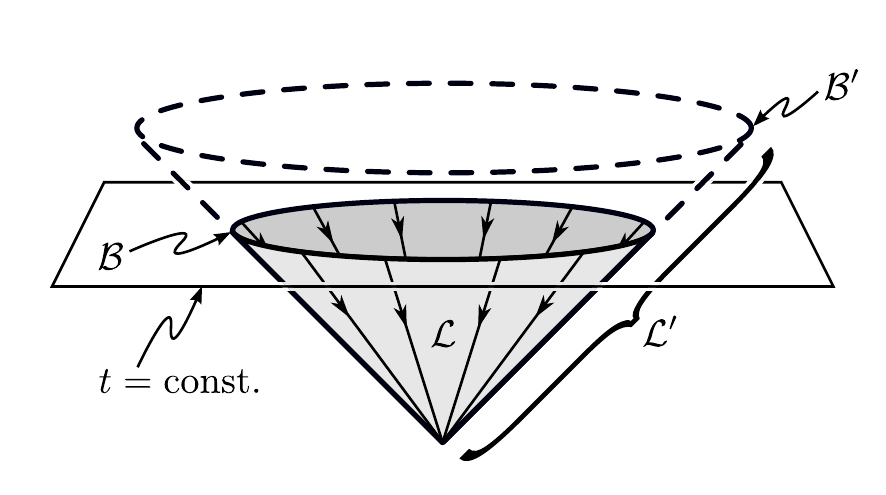}
\caption{Here we depict the past-directed lightsheet $\mathcal{L}$ of a simultaneous spherical surface $\mathcal{B}$, within an expanding FRW universe (\ref{FRWmetric}). If $\mathcal{B}$ is sufficiently small, we can expand the hologram $(\mathcal{B},\mathcal{L}$) by extending the converging null geodesics of $\mathcal{L}$ \emph{backwards} through $\mathcal{B}$. (For a past-directed $\mathcal{L}$, this extends the lightsheet towards the future.) This produces a new hologram $(\mathcal{B}',\mathcal{L}')$ that is a strict superset of the former: $\mathcal{L}'\supset \mathcal{L}$.  The new hologram must be considered the more fundamental description, as it contains all the information of the original hologram, and more besides. This process of \emph{backwards extension} can continue until  cosmological constraints intervene. The results of this maximisation procedure define the natural holograms to cover the FRW spacetime.}\label{hologrow}
\centering
\end{figure}

As illustrated in figure \ref{hologrow}, a superset hologram $(\mathcal{B}',\mathcal{L}')$ can be constructed from a (sufficiently small) candidate $(\mathcal{B},\mathcal{L})$  by extending the lightsheet $\mathcal{L}$ backwards through $\mathcal{B}$. If at some point this process fails, then $(\mathcal{B},\mathcal{L})$ will be maximal, and suitable for our universal covering. Indeed, there are two fundamental constraints that can cause backwards extension to fail:
\begin{enumerate}
\item \emph{The Geometric Constraint:} By definition, $\mathcal{L}$ is composed of null geodesics with \emph{nonpositive} expansion. This stipulation is a local representation of the notion that $\mathcal{L}$ should point ``inwards'' from $\mathcal{B}$, a key property that allowed Bousso to formulate his entropy bound (\ref{Bbound}) in the first place \cite{Bousso99a}. Backwards extension will therefore fail if we ever have $A[\mathcal{B}']< A[\mathcal{B}]$: the null rays from $\mathcal{B}'$ to $\mathcal{B}$ must then have \emph{positive} expansion, so $\mathcal{L}'$ will fail to be a valid lightsheet. 
\item \emph{The Causal  Constraint:} We require each hologram $(\mathcal{B},\mathcal{L})$ to lie inside the past lightcone of some hypothetical observer. This constraint is imposed by \emph{black hole complementarity} \cite{Susskind93,Susskind94}, which prevents us from applying the laws of quantum mechanics to systems that can never be observed in their entirety.\footnote
{Without complementarity, the unitary formation and evaporation of a black hole \cite{Mathur09,Polchinski16,Marolf17,Unruh17} would violate the no-cloning theorem \cite{Wooters82}. Even if a firewall forms at the scrambling time \cite{Almheiri13}, we still need complementarity to prevent cloning before then \cite{Susskind12,Bousso13}. A stricter interpretation of complementary would require $(\mathcal{B},\mathcal{L})$ to lie inside a \emph{causal diamond}, i.e.\ the intersection of some past lightcone and some future lightcone \cite{Bousso00, Bousso12}. We adopt the more tolerant version for now; in any case, this distinction would only be important in the very early universe (i.e.\ during inflation) when the particle horizon is closer than the event horizon.} While it is conceivable that the entropy bound (\ref{Bbound}) remains valid for lightsheets that break this constraint, these $\mathcal{L}$ cannot be be treated as quantum systems. Without a Hilbert space $\mathcal{H}_\mathcal{L}$ with known information capacity (\ref{InfoSat}) we cannot apply the quantum theory of section \ref{DoFs}. 
\end{enumerate}

\begin{figure*}[t]\centering
\includegraphics[scale=.938]{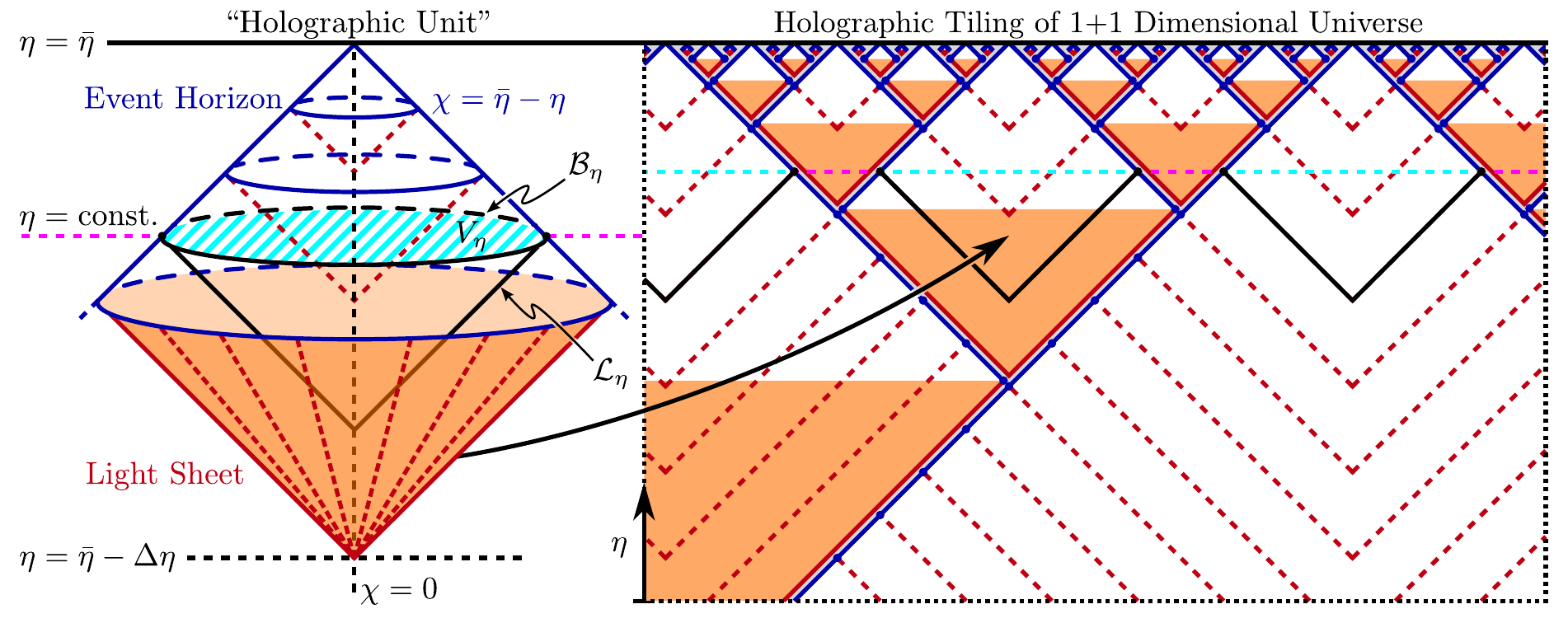}\vspace{-.25cm}
\caption{Holographic units are spherically symmetric causal diamonds, bounded into the future by a cosmological event horizon, and foliated by the past-directed lightsheets of the event horizon at each conformal time $\eta$. On the right, these units are arranged into a self-similar pattern that perfectly tiles an expanding universe with one spatial dimension and final conformal time $\bar{\eta}$. (We generalise this pattern to $D$ spatial dimensions in section \ref{Filling}.) Each holographic unit begins at $\eta=\bar \eta - 2^n \Delta \eta$ for some $n \in\mathbb{Z}$; all reference to the arbitrary scale $\Delta \eta$ can be removed by a natural averaging procedure described in the main text. On each spatial slice $\eta=\text{const,}$ the event horizon is a sphere $\mathcal{B}_\eta$ of area $A[\mathcal{B}_\eta]=\mathscr{A}(\bar{\eta}-\eta)[a(\eta)]^2$ that encloses a volume $V_\eta \equiv\mathscr{V}(\bar{\eta}-\eta)[a(\eta)]^3$; each $\mathcal{B}_\eta$ generates a past-directed lightsheet $\mathcal{L}_\eta$ with information capacity set by the holographic formula (\ref{InfoSat}). Note that even though the pattern covers the entire 1+1 dimensional spacetime without gaps or overlap, the (cyan shaded) volumes $V_\eta$  do not fill each spatial slice: some parts of the slice (magenta dashed line) are occupied by the lower half of a holographic unit (orange triangle) the information capacity of which will be counted on a future slice. Hence the number of spheres $\mathcal{B}_\eta$ in a large volume $V_*$ is $\mathscr{N}_*=\mu V_*/V_\eta$, for some ``filling factor'' $\mu\lesssim 1$.}\label{covering}
\centering
\end{figure*}

In a universe such as ours, which is expanding $\dot{a}>0$ and has low spatial curvature, holograms $(\mathcal{B},\mathcal{L})$ with \emph{past-directed} lightsheets $\mathcal{L}$ will always satisfy the geometric constraint. However, the causal constraint will halt backwards extension as soon as $\mathcal{B}$ coincides with the cosmological event horizon. In other words, a maximal past-directed hologram, centred at $\chi=0$, will have its boundary at
\begin{align}\label{pastB}
\mathcal{B}_\eta:\quad  \chi = \bar{\eta}- \eta,
\end{align}
where $\eta$ is the conformal time (\ref{etadef}) and 
\begin{align}\label{etabardef}
\lim_{\eta \to \bar{\eta}} a(\eta)=\infty
\end{align}
defines the final conformal time $\bar{\eta}$. (We check that $\bar{\eta}$ exists in section \ref{etabarcheck}.) Even if spatial curvature is large, the only way  (\ref{pastB}) will break down is if the universe is closed and the event horizon lies beyond the equator: $\bar{\eta}-\eta> \pi/2$. Then the geometric constraint can halt backwards extension before the event horizon is reached. However, $\bar{\eta}-\eta> \pi/2$ can only occur at very early times (during inflation) so we can ignore this special case for now. (We will revisit this issue in a separate publication, when we investigate quantum bias in the very early universe.) Of course, maximal holograms need not be centred on $\chi=0$; but if we place one hologram $(\mathcal{B}_\eta, \mathcal{L}_\eta)$ there, then a neighbouring maximal hologram $(\mathcal{B}_{\eta + \delta \eta}, \mathcal{L}_{\eta+\delta\eta})$ will have to also be centred at $\chi=0$ if the two are to be disjoint. In this fashion, maximal past-directed holograms naturally stack to form a spherically symmetric causal diamond, as depicted on the left of figure \ref{covering}. We will build our universal covering from these \emph{holographic units} in the next section. 

Before then, we should also consider \emph{future-directed} holograms. In contrast to the previous case, the causal constraint is unable to halt backwards extension, because if $(\mathcal{B},\mathcal{L})$ fits inside the event horizon, then $(\mathcal{B}',\mathcal{L}')$ will fit inside also. Instead, extension halts once $\mathcal{B}$ coincides with the apparent horizon,
\begin{align}\label{futureB}
 r_k(\chi_\mathrm{AH})= \left(\frac{1}{a^2}\left(\frac{\ud a}{\ud \eta}\right)^2 + k\right)^{-1/2},
\end{align}
by virtue of the geometric constraint. These holograms are unsuitable for our universal covering, for two distinct reasons. Firstly, the area of the apparent horizon (\ref{futureB}) clearly depends on $\ud a/\ud \eta $, so we would arrive at an information capacity $\mathcal{S}=\mathcal{S}(a,\ud a/\ud \eta)$ that is incompatible with the formula (\ref{VofS}) for quantum bias.\footnote{The theory summarised in appendix \ref{BiasDeriv} is valid for the general class $\mathcal{S}=\mathcal{S}(a,\int^\eta \ud \eta' f(a))$ \cite{Butcher18b}. It is doubtful whether these results can be generalised to $\mathcal{S}(a,\ud a/\ud \eta)$, as this form of information capacity requires a phase space that is not a cotangent bundle.} Secondly, the apparent horizon (\ref{futureB}) is determined by the behaviour of the scale factor, so any pattern of future-directed maximal holograms, intended to cover the universe with minimal overlap, will only succeed for a specific expansion history $a(\eta)$. This poses a serious problem for our approach, because $\mathcal{S}$ must be robust to arbitrary variations $\delta a(\eta)$ in order to be included in the semiclassical action $\mathcal{J}[a(\eta)]$.\footnote{Conceivably, there might be a general algorithm for covering spacetime with these holograms (with minimal overlap) valid for any $a(\eta)$; however, this would presumably define a \emph{non-local} functional $\mathcal{S}[a(\eta)]$ that would greatly exacerbate our first issue.} For the sake of practicality and generality, then, we must build our covering using the past-directed holographic units described in the previous paragraph.

\subsection{Holographic Covering}\label{Tiling}
If the classical action (\ref{IGa}) were an integral over a single causal diamond, then the holographic unit (on the left of fig.\ \ref{covering}) would provide all the structure we need. However, to make contact with the quantum theory of section \ref{DoFs}, it was necessary to integrate over a region (\ref{Mdef}) of fixed comoving volume, with a view to sending $\chi_*\to\infty$ at the end of our calculation. In order to count all the degrees of freedom in the action, we therefore need a systematic way to cover the entire FRW spacetime (\ref{FRWmetric}) with holographic units, such that there is minimal double counting from overlapping holograms.  In 1+1 dimensions, this problem has a particularly elegant solution, shown on the right of figure \ref{covering}. This two-dimensional picture will suffice to understand the calculation below, deriving the cosmological information capacity up to a numerical constant $\bar{\mu}$. Then, in the final section of this appendix, we will generalise this self-similar pattern to 3+1 dimensions, account for the small gaps or overlaps that arise, and determine the value of $\bar{\mu}$.

With a prototypical holographic covering at hand (fig.\ \ref{covering}) we aim to calculate the information capacity of some spatial slice $\eta=\text{const}$, within the integration region $\chi \in [0,\chi_*]$. We think of the bulk spacetime as composed of holograms $(\mathcal{B}_\eta,\mathcal{L}_\eta)$, with the state of each lightsheet $\mathcal{L}_\eta$ specified by information on the boundary $\mathcal{B}_\eta$. Hence, the information capacity on $\eta=\text{const,}$ is simply the information capacity (\ref{InfoBound}) of each sphere $\mathcal{B}_\eta$, multiplied by the number of these spheres $\mathscr{N}_*(\eta)$ within $\chi \in [0,\chi_*]$:
\begin{align}\label{Scounting}
\mathcal{S}= \mathscr{N}_* (\eta) \cdot \frac{A[\mathcal{B}_\eta]}{4\ell_\mathrm{pl}^2}.
\end{align}
If the spheres could be packed perfectly, without gap or overlap, then one might expect 
\begin{align}\label{Nguess}
\mathscr{N}_*(\eta)\stackrel{?}{=} \frac{V_*}{V_\eta},
\end{align}
where $V_*=\mathscr{V}(\chi_*) [a(\eta)]^3$ is the volume of the integration region $\chi \in [0,\chi_*]$, and $V_\eta=\mathscr{V}(\bar{\eta}-\eta)[a(\eta)]^3$ is the volume enclosed by each $\mathcal{B}_\eta$. However, figure \ref{covering} shows us that this is not the case. Even for the 1+1 dimensional tiling, which does indeed cover the universe without gaps or overlap, the $\mathcal{B}_\eta$ do not fill each spatial slice. In general, only a fraction 
\begin{align}
\mu \equiv \frac{\mathscr{N}_*(\eta)V_\eta}{V_*}\lesssim1
\end{align}
of the volume is taken up by the $\mathcal{B}_\eta$; the rest is occupied by the lower half of other (smaller) holographic units, foliated by holograms with their boundaries on future slices. 

Consulting figure \ref{covering}, it appears that $\mu$ will oscillate -- decreasing from $\mu=1$, to $\mu =1/2$, as the spatial slice ascends through each cycle $\eta\in [\bar{\eta}- 2^n\Delta \eta , \bar{\eta}- 2^{n-1}\Delta \eta)$. However, the phase of this oscillation clearly depends on the arbitrary scale $\Delta \eta$:
\begin{align}
\mu&=\mu\!\left(\frac{\bar{\eta}-\eta}{\Delta \eta}\right).
\end{align}
Fortunately, there is a natural way to remove this spurious feature: a \emph{unique} average over $\Delta \eta$ that recovers the symmetry of the underlying spacetime. As we will soon show, this provides a physically well-defined \emph{constant} value
\begin{align}\label{mubardef}
\bar{\mu}\equiv \langle\mu \rangle_{\Delta \eta}=\frac{1}{\ln m}\int^{mx}_x  \mu\!\left(\frac{\bar{\eta}-\eta}{\Delta \eta}\right)\frac{\ud (\Delta \eta)}{\Delta \eta}
\end{align}
that correctly counts the spheres $\mathcal{B}_\eta$ in $\chi \in [0,\chi_*]$ without reference to $\Delta \eta$:
\begin{align}\label{Ncorrect}
\mathscr{N}_*(\eta)=\frac{\bar{\mu} V_*}{V_\eta} = \frac{\bar{\mu} \mathscr{V}_*}{\mathscr{V}(\bar{\eta}-\eta)}.
\end{align}
Inserting this well-defined counting into equation (\ref{Scounting}) we finally obtain the information capacity 
\begin{align}\label{Sderived}
\mathcal{S}=\frac{\bar{\mu}\mathscr{V}_*}{\mathscr{V}(\bar{\eta}-\eta)} \cdot \frac{\mathscr{A}(\bar{\eta}-\eta) [a(\eta)]^2}{4 \ell^2_\mathrm{pl}},
\end{align}
as used in the section \ref{HoloCap}. 

To finish this derivation, we must justify the averaging procedure (\ref{mubardef}) and show that it does not depend on the choice of $x>0$. To this end, let us consider an arbitrary function $f$ that (like $\mu$) depends only on the phase of a self-similar holographic covering at conformal time $\eta$. As such, $f$ will have the following structure:
\begin{align}\label{fsimilar}
f&=f\!\left(\frac{\bar{\eta}-\eta}{\Delta \eta}\right), & f(mx)&=f(x),\quad \forall\ x>0,
\end{align}
where $m\in\{2,3,\ldots\}$ is the scaling-factor under which the pattern is self-similar. (The pattern in figure \ref{covering} has $m=2$.) For a function with these properties, any arithmetic mean over $\Delta \eta$ can be represented as an integral over a single scaling cycle:
\begin{align}\label{genmean}
\langle f \rangle_{\Delta\eta}\equiv \int^{mx}_x  f\!\left(\frac{\bar{\eta}-\eta}{\Delta \eta}\right) g(\Delta \eta)  \ud (\Delta \eta),
\end{align}
with some measure $g(\Delta \eta)$ normalised by 
\begin{align}\label{gnorm}
\int^{mx}_x g(\Delta \eta)  \ud (\Delta \eta)=1.
\end{align}
We will seek a $g(\Delta \eta)$ that allows $\langle f \rangle_{\Delta\eta}$ to respect the symmetry of the underlying spacetime, for every $f$ with the appropriate structure (\ref{fsimilar}).

Let us assume for the moment that $k=0$, so that the underlying spacetime has the metric
\begin{align}\label{FRWsimple}
\ud s^2 = [a(\eta)]^2\left(-\ud \eta^2+ \ud \chi^2 + \chi^2\ud \Omega^2\right).
\end{align}
Note that this spacetime is invariant under the following conformal transformation:
\begin{align}\label{Weyl}
\ud s^2 &\to \left(\frac{a\!\left(\alpha \eta + (1-\alpha) \bar{\eta}\right)}{ a(\eta)}\right)^2 \alpha ^2\ud s^2, 
\end{align}
for any constant $\alpha>0$; indeed, the above transformation is equivalent to a coordinate rescaling,
\begin{align}\label{rescale}
\eta&\to \alpha \eta + (1-\alpha) \bar{\eta},& \chi &\to \alpha \chi,
\end{align}
that leaves $\bar{\eta}$ invariant. We notice, however, that the holographic covering will break this symmetry almost entirely -- all that survives are transformations with $\alpha \in \{m^n: n\in \mathbb{Z}\}$. As a case in point, consider $f$. Because this is purely a function of  the phase of the holographic covering, it will not depend on the scale factor, and so is invariant under the Weyl transformation (\ref{Weyl}). If this function were to respect the full symmetry of the underlying spacetime, it would therefore also need to be invariant under the coordinate rescaling (\ref{rescale}). However, its properties (\ref{fsimilar}) only guarantee invariance for $\alpha = m^n$, $n \in \mathbb{Z}$. 

Now, by construction, the average (\ref{genmean}) is also independent of $a(\eta)$, and hence invariant under the Weyl transformation (\ref{Weyl}). Thus, $\langle f \rangle_{\Delta\eta}$ will recover the full symmetry of the underlying spacetime (\ref{FRWsimple}) if and only if it is invariant under the coordinate rescaling (\ref{rescale}) for all $\alpha >0$. In other words, $\langle f \rangle_{\Delta\eta}$ cannot depend on $\eta$ at all. Thus we seek a measure $g(\Delta \eta)$ that ensures
\begin{align}\label{meanconst}
\langle f \rangle_{\Delta\eta}= \text{const,}
\end{align}
for all $f$ with the aforementioned properties (\ref{fsimilar}). But note that
\begin{align}\label{meanparts}
 \partial_\eta \langle f \rangle_{\Delta\eta}&= \int^{mx}_x  \partial_\eta f\!\left(\frac{\bar{\eta}-\eta}{\Delta \eta}\right) g(\Delta \eta) \ud (\Delta \eta)
\\\nonumber
&=\int^{mx}_x \left( \frac{\Delta \eta}{\bar{\eta}-\eta} \right)\partial_{\Delta\eta} f\!\left(\frac{\bar{\eta}-\eta}{\Delta \eta}\right) g(\Delta \eta) \ud (\Delta \eta)
\\\nonumber
&=\frac{1}{(\bar{\eta}-\eta)}\bigg\{\left[ f\!\left(\frac{\bar{\eta}-\eta}{\Delta \eta}\right) g(\Delta \eta) \Delta \eta \right]^{mx}_{x}
\\\nonumber
&\quad {} - \int^{mx}_x f\!\left(\frac{\bar{\eta}-\eta}{\Delta \eta}\right)  \partial_{\Delta\eta} \Big( g(\Delta \eta) \Delta \eta\Big) \ud (\Delta \eta) \bigg\}.
\end{align}
Hence the symmetry condition (\ref{meanconst}) requires this last line to vanish for every $f$ obeying (\ref{fsimilar}). This will happen if and only if
\begin{align}
 \partial_{\Delta\eta} \Big( g(\Delta \eta) \Delta \eta\Big) =0,\qquad \forall\ \Delta \eta \in [x,mx],
\end{align}
and recalling the normalisation (\ref{gnorm}) we see that
\begin{align}
g(\Delta \eta) = \frac{1}{\ln m}\cdot\frac{1}{\Delta \eta}, \qquad \forall\ \Delta \eta \in [x,mx],
\end{align}
is the only solution. Thus the \emph{unique} mean (\ref{genmean}) that recovers the symmetry of the underlying spacetime is
\begin{align}\label{fbardef}
\langle f \rangle_{\Delta\eta}&\equiv \frac{1}{\ln m}\int^{mx}_x  f\!\left(\frac{\bar{\eta}-\eta}{\Delta \eta}\right) \frac{\ud (\Delta \eta)}{\Delta \eta},
\end{align}
as used in equation (\ref{mubardef}). Furthermore, it is easy to check that this construction does not depend on our choice of $x$:
\begin{align}\nonumber
\partial_x \langle f \rangle_{\Delta\eta}&= \frac{1}{\ln m}\left[m\cdot\frac{1}{mx}f\!\left(\frac{\bar{\eta}-\eta}{mx}\right) - \frac{1}{x}f\!\left(\frac{\bar{\eta}-\eta}{x}\right)\right]\\
&=0,
\end{align}
by virtue of the second property (\ref{fsimilar}). 

For $k=\pm 1$, the holographic covering will not be exactly self-similar (spatial curvature introduces a special comoving scale $\chi=1$) and the Weyl transformation (\ref{Weyl}) will not be an exact symmetry. Nonetheless, when the event horizon is much smaller than the radius of spatial curvature $|k|(\bar{\eta}-\eta) \ll 1$, the $k=0$ case will be an excellent approximation, and we can safely use the average (\ref{fbardef}) to define $\bar{\mu}$. This approximation can only break down in the very early universe.

\begin{figure}
    \centering
\includegraphics[width=0.46\textwidth]{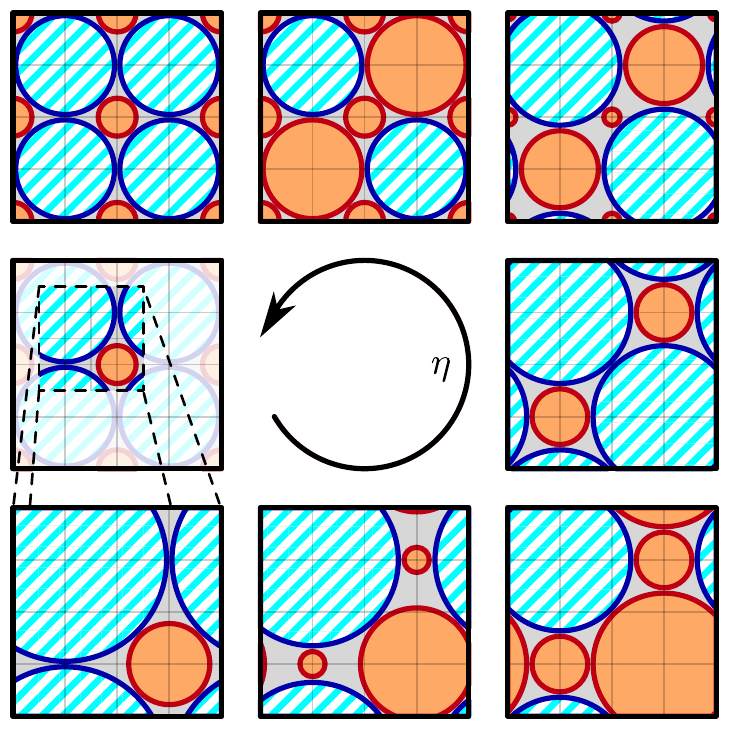}
\caption{The cycle above generalises the self-similar pattern of figure \ref{covering}, packing holographic units into an expanding 2+1 dimensional universe without overlap. Each frame  represents the state of a comoving square lattice on a sequence of spatial slices $\eta=\text{const}$. The pattern is easiest follow in reverse chronological order (clockwise) starting from the top-left frame: as $\eta$ decreases, the comoving radii of the event horizons (blue circles) grow, while the initial lightsheets (red circles) shrink. Whenever two event horizons touch (frames 1 and 4) every other horizon is transformed into an initial lightsheet (frames 2 and 5). These transitions represent the ``corner'' of a holographic unit, e.g.\ the $\eta=\bar{\eta}-\Delta \eta/2$ slice of the unit depicted on the left of figure \ref{covering}. This process prevents any holographic unit from overlapping, but allows small gaps (grey) to appear in the covering. Once we reach the bottom-left frame, the lattice has returned to its starting state, scaled up by a factor $m=2$. This algorithm is easily generalised  to pack holographic units in $D$+1 dimensions, or modified to construct (partially overlapping) patterns that cover the entire spacetime.}\label{2dholo}
\end{figure}

\subsection{Filling Factor}\label{Filling}
It is presumably impossible to generalise figure \ref{covering} to 3+1 dimensions without introducing either gaps (regions not covered by a holographic unit) or overlaps (regions covered by more than one unit). Nonetheless, we can aim to make these defects as small as possible, and correct for the resultant under/overcounting when we calculate the filling factor $\bar{\mu}$. 

For instance, suppose we construct a reasonably efficient \emph{packing} pattern, with small gaps but no overlap, as described in figure \ref{2dholo}. Some volume-fraction $\mu$ of each spatial slice will be covered by the $\mathcal{B}_\eta$, i.e.\ the (cyan shaded) horizon-bound regions that form the top half of each holographic unit; also, some fraction $\nu$ will be covered by the (orange) lightsheet-bound regions that constitute the bottom half of each unit. The tiling of figure \ref{covering} had perfect coverage $\mu+\nu=1$ on every slice, so we were able to identify $\bar{\mu}=\langle \mu \rangle_{\Delta\eta}$ using the invariant average (\ref{fbardef}). However, the gaps $\mu +\nu <1$ in figure \ref{2dholo} mean that parts of the spacetime are not described by any hologram ($\mathcal{B}_\eta,\mathcal{L}_\eta)$; as such, $\langle \mu \rangle_{\Delta\eta}$ for this pattern will inevitably underestimate $\bar{\mu}$, and only provide a lower bound on $\mathcal{S}$. Conversely, a reasonably efficient \emph{covering}, with overlaps but no gaps ($\mu+\nu>1$) will yield a $\langle \mu \rangle_{\Delta\eta}$ that slightly overestimates $\bar{\mu}$, due to double counting. To correct for these defects, we identify 
\begin{align}\label{mubargen}
\bar{\mu}\equiv\frac{\langle \mu \rangle_{\Delta\eta}}{\langle \mu +\nu \rangle_{\Delta\eta} }.
\end{align}
This formula generalises the earlier definition (\ref{mubardef}), accounting for any net deficit $\langle \mu +\nu \rangle_{\Delta\eta}<1$ (due to gaps) or excess $\langle \mu +\nu \rangle_{\Delta\eta}>1$ (due to overlap) in the holographic coverage. Crucially, this formula is \emph{completely independent} of our choice of holographic pattern. We can evaluate the right-hand side of equation (\ref{mubargen}) using any self-similar configuration -- the value of $\bar{\mu}$ will be exactly the same. As a consequence, there is no need to worry about finding a \emph{maximally efficient} packing or covering. Finding a more efficient pattern will simply move $\langle \mu +\nu \rangle_{\Delta\eta}$ closer to 1, and $\langle \mu \rangle_{\Delta\eta}$ closer to $\bar{\mu}$, with $\bar{\mu}=\langle \mu \rangle_{\Delta\eta}/\langle \mu +\nu \rangle_{\Delta\eta}$ unchanged. (In other words, $\bar{\mu}$ is the limiting value of $\langle \mu \rangle_{\Delta\eta}$ as the pattern is made more efficient.) To prove this surprising fact, and determine $\bar{\mu}$ numerically, we now describe a completely general self-similar pattern of holographic units.

Let us consider a spatially flat FRW universe with $D+1$ dimensions, and introduce a pattern of holographic units that are self-similar under a rescaling $\bar{\eta}-\eta \to m(\bar{\eta}-\eta)$ for some $m\in\{2,3,\ldots\}$. To fully describe any such pattern, we need only specify its behaviour within  a single scaling cycle:
\begin{align}\label{cycle}
\eta =\bar{\eta}- s \Delta \eta,\qquad s\in[1,m),
\end{align}
where $\Delta \eta$ is an arbitrary scale that will need to be averaged out (\ref{fbardef}) at the end of the calculation. As we saw in figure \ref{2dholo}, each holographic unit will contain two types of spatial region: (i) the (cyan shaded) sheres bound by a cosmological event horizon (blue circle);  and (ii) the (orange) spheres bound by an initial lightsheet (red circle). If we imagine the spatial sections $\eta=\bar{\eta}- s \Delta \eta$ of our generic pattern, and increase $s$ through $s\in [1,m)$, the comoving radii of the horizon-bound spheres will grow according to $\chi=\bar{\eta}-\eta=s\Delta \eta$, while the radii of lightsheet-bound spheres will shrink at the same rate, until they vanish entirely. In addition, there will be particular phases of the pattern $s_i\in(1,m)$ where some holographic units have \emph{corners}: a subset of the horizon-bound spheres will suddenly transform into lightsheet-bound spheres. (To avoid ambiguity, any transitions at $s=1$ should be considered to happen at $s=1+\epsilon$, for some small $\epsilon>0$.)

\begin{figure}
    \centering
\includegraphics[width=0.49\textwidth]{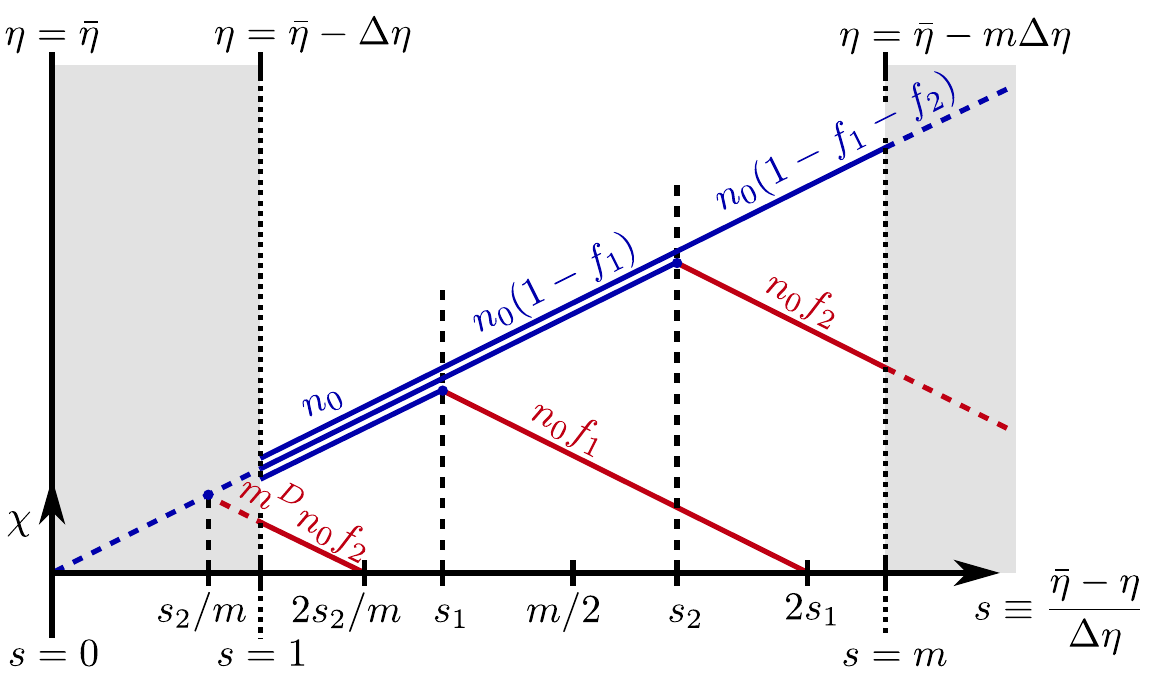}
\vspace{-0.6cm}
\caption{Over a single scaling cycle (\ref{cycle}) the spatial slices of a self-similar pattern of holographic units undergo two types of evolution. \emph{Continuous}: as $s$ increases, the comoving radii of the event horizons (blue) grow, while the initial lightsheets (red) shrink. \emph{Discrete}: at each $s=s_i$, a fraction of the holographic units have corner transitions -- their event horizons terminate and become initial lightsheets. The diagram above represents a simple example, with two transitions: $s_1<m/2<s_2$. The terms running along diagonal lines indicate the number such spheres within the integration region $\chi \in [0,\chi_*]$. Note that the $s_2$ transition produces $n_0 f_2$ lightsheet-bound spheres which still exist  at end of the cycle $s=m$. By the self-similarity of the pattern, there must be $ m^D \times ( n_0 f_2 )$ similar spheres (smaller by a factor of $1/m$) that survive the previous cycle $s\in[1/m,1)$ and enter the current cycle at $s=1$.}\label{ScaleDiag}
\end{figure}

Figure \ref{ScaleDiag} illustrates how the number and scale of each type of sphere will evolve over the cycle (\ref{cycle}). At $s=1$, we have some number
\begin{align}\label{n0def}
n_0\equiv \mathscr{N}_*|_{\eta=\bar{\eta}-\Delta\eta}\propto \mathscr{V}_*/(\Delta \eta)^D
\end{align}
of horizon-bound spheres within the integration region $\chi\in[0,\chi_*]$. As we increase $s$, we encounter each transition $s=s_i$ in turn, with $n_0 f_i$ horizon-bound spheres becoming lightsheet-bound spheres. Consequently, the horizon-bound spheres occupy a volume-fraction 
\begin{align}\nonumber
\mu(s)& \equiv\left. \frac{\mathscr{N}_*(\eta) V_\eta}{V_*}\right|_{\eta=\bar{\eta}-s\Delta \eta}\\ \nonumber
&=  \frac{\mathscr{V}(s\Delta \eta)}{\mathscr{V}_*}\mathscr{N}_*(\bar{\eta}-s \Delta \eta)\\ \nonumber
&=\frac{\mathscr{V}(1)(s\Delta \eta)^D }{\mathscr{V}_*}\left[n_0- \sum_i n_0 f_i H(s-s_i)\right]\\\label{mus}
&=\mu(1)s^D\left[1- \sum_i  f_i H(s-s_i)\right],
\end{align}
where $H$ is the Heaviside step function and 
\begin{align}
\mu(1)= \frac{\mathscr{V}(1)(\Delta \eta)^D n_0}{\mathscr{V}_*}
\end{align}
is a numerical constant.\footnote{$\mathscr{V}(1)=\pi^{D/2}/\Gamma(1+ D/2)$ is the volume enclosed by a unit sphere in $D$ dimensions. Consulting equation (\ref{n0def}) we see that $\mu(1)$ is independent of the scale $\Delta \eta$ and the integration volume $\mathscr{V}_*$.} Equation (\ref{mus}) was derived for the cycle $s\in[1,m)$, but must continue to hold at $s=m$ because there are no transitions at $s=m$. Hence, the self-similarity (\ref{fsimilar}) of the pattern implies
\begin{align}\label{sumf}
\mu(1)&=\mu(m) &\Rightarrow\quad  \sum_i  f_i&=1-m^{-D}.
\end{align}
In addition to the volume fraction of horizon-bound spheres (\ref{mus}), we must now account for the lightsheet-bound spheres. 

Consulting figure \ref{ScaleDiag} again, we see that the $n_0f_i$ lightsheet-bound spheres that form at $s=s_i$ have radius  $\chi_{i}=(2s_i-s)\Delta \eta$ and vanish at $s=2s_i$. Those that appear at $s_i>m/2$ will still exist at the end of the cycle: $s=m\ \Rightarrow\ \chi_{i}=(2s_i-m)\Delta \eta>0$. Hence $m^Dn_0f_i$ lightsheet-bound spheres, of radius $\chi'_{i}= ((2s_i/m)-s)\Delta \eta$, must have survived the previous cycle $s\in[1/m,1)$. We conclude that the volume-fraction of lightsheet-bound spheres is 
\begin{widetext}
\begin{align}\nonumber
\nu(s) &= \frac{1}{\mathscr{V}_*}\left[\sum_i n_0 f_i \mathscr{V}(\chi_{i}) H(s-s_i)H(2s_i-s)+ \sum_{i: s_i>m/2}\!\! m^D n_0 f_i \mathscr{V}(\chi'_{i})H\bigg(\frac{2s_i}{m}-s\bigg)\! \right] \\\label{nus}
&=\mu(1) \left[\sum_i f_i\left(2s_i-s\right)^D H(s-s_i)H(2s_i-s) +  \sum_{i: s_i>m/2}\!\! f_i \left(2s_i-ms\right)^D H\bigg(\frac{2s_i}{m}-s\bigg)\right].
\end{align}
Although this equation was only derived for $s\in [1,m)$, it must also hold at $s=m$ by continuity. In contrast to the previous result (\ref{mus}), equation (\ref{nus}) is automatically self-similar: $\nu(1)=\nu(m)$; hence we obtain no constraints on the $f_i$ besides equation (\ref{sumf}).

To recover the symmetry of the underlying spacetime, and obtain the invariant versions of $\mu$ and $\nu$, we now average over the arbitrary scale $\Delta \eta$. With $\bar{\eta}$ and $\eta$ fixed, equation (\ref{cycle}) implies that the natural average (\ref{fbardef}) can be written as follows:
\begin{align}\label{smean}
\langle f \rangle_{\Delta\eta}&= \frac{1}{\ln m}\int^{m}_1  f(s) \frac{\ud s}{ s},
\end{align}
where we have chosen $x=(\bar{\eta}-\eta)/m$ to align this integral with the cycle $s\in[1,m)$. Taking the average of equation (\ref{mus})  we obtain
\begin{align}\label{muav}
\langle \mu \rangle_{\Delta\eta}= \frac{ \mu(1)}{\ln m}\int^{m}_1   s^{D}\left[1- \sum_i  f_i H(s-s_i)\right] \frac{\ud s}{ s} = \frac{\mu(1)}{D\ln m}\left(m^D - 1 - \sum_i f_i (m^D-s_i^D)\right)=\frac{\mu(1)}{D\ln m}\sum_i f_i s_i^D,
\end{align}
where equation (\ref{sumf}) was used for the last step. Next, we take the average of equation (\ref{nus}):
\begin{align}
\langle \nu \rangle_{\Delta\eta}&=\frac{\mu(1)}{\ln m}\left[\sum_{i:s_i\le m/2}\!\!f_i\!\int_{s_i}^{2s_i}\frac{(2s_i-s)^D\ud s}{s}+\!\! \sum_{i:s_i> m/2}\!\!f_i\!\int_{s_i}^{m}\frac{(2s_i-s)^D\ud s}{s}+ \!\!\sum_{i:s_i> m/2}\!\!f_i\!\int_{1}^{2s_i/m}\frac{(2s_i-m s)^D\ud s}{s}\right].
\end{align}
\end{widetext}
Rescaling $s\to s/m$ in the third set of integrals, this simplifies to
\begin{align}\nonumber
\langle \nu \rangle_{\Delta\eta}&= \frac{\mu(1)}{\ln m}\sum_{i}f_i\int_{s_i}^{2s_i}\frac{(2s_i-s)^D\ud s}{s}\\
&= \frac{\mu(1)}{\ln m} \left(\sum_{i}f_i s_i^D\right) \int_{1}^{2}\frac{(2-s)^D\ud s}{s},
\end{align}
where we replaced dummy variables $s \to s_i s$ to produce the final line. We conclude that the invariant coverage is 
\begin{align}\label{covav}
\langle\mu+ \nu\rangle_{\Delta\eta}&=\langle\mu\rangle_{\Delta\eta}+\langle \nu\rangle_{\Delta\eta}\\ \nonumber
&= \frac{\mu(1)}{\ln m} \left[\frac{1}{D}+ \int_{1}^{2}\frac{(2-s)^D\ud s}{s}\right]\sum_{i}f_i s_i^D,
\end{align}
for a general self-similar pattern of holographic units.

We now have everything needed to calculate the filling factor (\ref{mubargen}). Dividing equation (\ref{muav}) by equation (\ref{covav}), we obtain our final result:
\begin{align}\label{mubarD}
\bar{\mu}= \left(1+D \cdot  \int_{1}^{2}\frac{ (2-s)^D\ud s}{s}\right)^{-1}.
\end{align}
Remarkably, all the variables $\{m,s_i,f_i,\mu(1)\}$ have cancelled, so the details of the pattern are completely irrelevant. This demonstrates the naturalness of our definition (\ref{mubargen}) and provides an extremely simple formula for $\bar{\mu}$. For our universe, with $D=3$ spatial dimensions, the holographic filling factor (\ref{mubarD}) is simply
\begin{align}\label{mubar3}
\bar{\mu}=\frac{1}{24 \ln 2 -15}= 0.61142\ldots
\end{align}
This completes our calculation of the cosmological holographic information capacity (\ref{Sderived}).

\newpage

\bibliography{ShCDM}

\end{document}